\documentclass[trackchanges, twocolumn]{aastex701}

\usepackage{bm}
\usepackage{array}
\usepackage{amsmath}

\shorttitle{Suppressing star formation in bulge-dominated early-type galaxies}
\shortauthors{L.~E.~Porter et al.}

\begin{document}

\defcitealias{Jeffreson2024b}{J24}

\title{What Suppresses Star Formation in Bulge-Dominated Early-Type Galaxies?}

\author[orcid=0000-0002-7795-2267,sname='Porter']{Lori E. Porter}
\affiliation{Department of Astronomy, Columbia University, 550 W. 120th Street, New York, NY, 10027, USA}
\email{lep2176@columbia.edu}  

\author[orcid=0000-0002-4232-0200, sname='Jeffreson']{S. M. R. Jeffreson} 
\affiliation{Center for Astrophysics, Harvard \& Smithsonian, 60 Garden Street, Cambridge, MA 02138, USA}
\email{sarah.jeffreson@cfa.harvard.edu}

\author[orcid=0000-0003-2630-9228, sname=Bryan]{Greg L. Bryan}
\affiliation{Department of Astronomy, Columbia University, 550 W. 120th Street, New York, NY, 10027, USA}
\email{gb2141@columbia.edu}

\author[orcid=0000-0001-6950-1629, sname=Hernquist]{Lars Hernquist}
\affiliation{Center for Astrophysics, Harvard \& Smithsonian, 60 Garden Street, Cambridge, MA 02138, USA}
\email{lhernquist@cfa.harvard.edu}



\begin{abstract}
We investigate the physical origin of star formation suppression in gas-rich early-type galaxies using five high-resolution hydrodynamical idealized galaxy simulations, performed with the moving-mesh code {\sc AREPO}. 
These simulations include one Milky Way-like galaxy and four early-type galaxies, of which one early-type galaxy is found to have significantly less star formation despite a substantial molecular gas reservoir. We apply a modified virial theorem to the overdensities in each galaxy to quantify the forces regulating their stability and thus star formation.
We find evidence that, in the suppressed galaxy, strong Coriolis forces driven by elevated galactic shear may inhibit gravitational collapse. This is caused by the galaxy's high central compactness, providing a physical mechanism for the suppression of star formation that does not require the removal of molecular gas. In contrast, less compact early-type galaxies host more gravity-dominated clouds and therefore exhibit higher star formation rates. However, we find that this gravitational stability occurs without significantly increasing the classical Toomre-$Q$ parameter, and therefore a  new criterion for suppressed star formation may be needed. We also discuss the impact of our choice of overdensity scale and connections to observations of molecular clouds. 
\end{abstract}



\section{Introduction} \label{Sec::Intro}

The pace and efficiency of star formation are key physical mechanisms that govern galaxy evolution. They determine galaxy stellar masses \citep{Feulner2005a, Furlong2015, Johnston2015, Salmon2015}, drive chemical enrichment (\citealt{Nomoto2013, Maiolino2019a} and references therein), power feedback \citep{Hopkins2014, Hayward2017, Burkhart2018}, and shape the interstellar medium (ISM; \citealt{deAvillez2005, Popping2014, Pettitt2017, Scoville2017, Saintonge2022}). As a result, galaxies with dramatically reduced or even halted (quenched) star formation represent a unique galactic population.

External quenching mechanisms such as ram-pressure stripping \citep{Book2010, Jaffe2015, Steinhauser2016, Simpson2018}, galaxy interactions \citep{Kennicutt1998, VanDenBosch2008, Kawata2008}, and AGN feedback \citep{Fabian2012, Dubois2013, Piotrowska2022, Bluck2023} can prevent star formation by physically removing a galaxy's gas. However, many quenched galaxies retain substantial cold gas reservoirs, implying the existence of a mechanism that reduces the levels of star formation in galaxies with relatively high gas fractions \citep{Martig2009, Martig2013, Davis2014, Liu2021}.

Elliptical early-type galaxies (ETGs) in the local Universe may reflect the influence of such mechanisms. Cold and molecular gas in nearby ETGs has been detected for decades \citep{Wiklind1986, Phillips1987, Welch2003, Combes2007, Young2011, Davis2019}, with more massive galaxies having some of the largest gas reservoirs \citep{Salome2005, Russell2016, OSullivan2018}. These galaxies also exhibit longer depletion times, suggesting lower star formation rates (SFRs) compared to the available gas supply. For example, quenched galaxies in the Sloan Digital Sky Survey (SDSS) have been observed to show star formation efficiencies (SFEs; $\dot M_{\star}/M_{\rm gas}$) two orders of magnitude lower than the star-forming main sequence \citep{Piotrowska2022}. \citet{Colombo2020} also find that diminishing a galaxy's gas, expressed through the molecular gas fraction, is not enough to explain the offset from the star-forming main sequence---there is also dependence on the SFE.

Instead, dynamical suppression (also referred to as `morphological' quenching, which places more emphasis on the presence of the central stellar component) may be a prominent quenching mechanism within the gas of bulge-dominated early-type galaxies \citep{Martig2009, Martig2013, Gensior2020, Gensior2021}. This occurs when an increased potential well from the stellar bulge causes an increase in angular velocity $\Omega$, velocity dispersion $\sigma$, and the Toomre-$Q$ stability criterion $Q$ ($Q=\kappa\sigma_{\rm R,gas}/\pi G\Sigma_{\rm gas}$, where $\kappa$ is the epicyclic frequency; \citealt{Toomre1964}), leaving the gas disk stable against collapse \citep{Martig2009}.

It is unclear to what extent this possible quenching mechanism could be operating in practice. Observations by \citet{Huang2014} find a weak relationship between stellar density and SFR, particularly when compared to the stronger relationship between depletion time and SFR. Some SDSS galaxies with bulge-dominated stellar distributions also show SFRs equivalent to the main sequence when total stellar mass is taken into account \citep{Belfiore2018}. This result is similar to \citet{Su2019}, where morphological quenching is not sustained within massive galaxies ($M_{\rm halo}\sim10^{12}-10^{14}\,{\rm M_{\odot}}$) and galaxy morphology has only a small effect on $Q$. In contrast, \citet{Kretschmer2020} reproduce quenching without AGN in a single cosmological zoom-in simulation, but only if the SFE per freefall time ($\epsilon_{\rm ff}$) is not constant, and is instead dependent on the gas virial parameter ($\alpha_{\rm vir}$) and Mach number ($\mathcal{M}$).  

\citet{Gensior2020} specifically test whether this strong bulge component directly causes quenching, finding that only a dynamics-dependent star formation prescription (reliant on $\alpha_{\rm vir}$) suppresses central star formation (where we define star formation suppression as falling well below the galactic star-forming main sequence). Without the presence of this central spheroidal component, galaxy-scale SFR varies little between models, and gas properties such as velocity dispersion and turbulent pressure only differ in the bulge-dominated regions. \citet{Gensior2021} suggest that dynamical suppression is stronger for lower gas fractions, and may regulate the baryon cycle to drive galaxies off the main sequence at redshifts of $z\leq1.4$ and masses $M_{\star}\approx3\times10^{10}\,{\rm M_{\odot}}$.

Furthermore, as described by \citet{Burger2025}, few hydrodynamical cosmological simulations are able to model the physics of small-scale star formation in a variety of galaxy populations. As a result, these simulations are less suited to studying quenching pathways \citep{Gensior2020, Piotrowska2022}. To create new cosmological simulations with sub-grid processes that can be modeled by physically motivated analytic theory, allowing for more comprehensive predictions of physics such as star formation quenching, the Learning the Universe collaboration introduced the GalactISM simulation suite in \citet{Jeffreson2024b}, henceforth referred to as \citetalias{Jeffreson2024b}. This suite consists of six high-resolution chemodynamical isolated galaxy simulations spanning the galactic main sequence and quenched populations, including four ETGs, a MW-like galaxy, and an NGC 300-like galaxy. 

From this suite, \citetalias{Jeffreson2024b} find that the relationship between midplane gas pressure and SFR surface density \citep{Ostriker2011, Ostriker2022} is an improved model for star formation compared to Kennicutt-Schmidt \citep{Kennicutt1998}.The equation of state between the gas density and pressure also varies strongly with epicyclic frequency ($\kappa^2=R^{-3}d(\Omega^2R^4)/dR$) when $\kappa$ is high, which is noted for future simulations. 

However, despite being initialized from the same observationally-motivated scaling relations, one ETG in \citetalias{Jeffreson2024b} shows a lower specific SFR (sSFR), referred to as the dynamically suppressed or `quenched' ETG. Of the ETGs, this galaxy has the highest galactic rotation ($\Omega$), $\kappa$, circular velocity ($v_{\rm circ}$), and the most compact stellar bulge. Yet, this galaxy does not show a significantly elevated velocity dispersion or Toomre-$Q$ parameter compared to the other ETGs, in contrast with results found by other studies of dynamical suppression \citep{Martig2009, Gensior2020}. A detailed investigation into the relationship of galactic properties with the suppression of star formation is clearly needed, as \citetalias{Jeffreson2024b} notes the onset of dynamical suppression is a regime not easily parameterized by either pressure-regulated star formation or by Kennicutt-Schmidt due to a nonlinear dependence on $\kappa$.

In this paper, we build on work by \citetalias{Jeffreson2024b} and investigate the cause of suppressed star formation within bulge-dominated ETGs through the lens of a modified virial theorem. The GalactISM simulation suite, with its high resolution and capacity to model star formation and stellar feedback in a multiphase ISM, provides a crucial framework for investigating star formation (or lack thereof). 

This paper is organized as follows: In Section~\ref{Sec::SimulationOverview} we review the simulation suite from \citetalias{Jeffreson2024b} and detail how overdensities (clouds) within the simulations are defined. In Section~\ref{Sec::VirialTheorem} we describe the modified virial theorem and how energy terms are calculated; the analysis of these energies is presented in Section~\ref{Sec::Results}. In Section~\ref{Sec::Discussion} we compare our work to previous simulations investigating star formation suppression (\ref{Sec::Discussion_PriorWork}), discuss recent results from GMC observations (\ref{Sec::Discussion_Observations}), review our definition of cloud scales and caveats (\ref{Sec::Discussion_Caveats}), and reiterate our primary conclusions and potential for future work in Section~\ref{Sec::conclusions}.

\section{Simulations}\label{Sec::SimulationOverview}

\begin{table*} 
\begin{centering}
\vspace{0.5cm}
  \begin{tabular}{c c c c c c c}
  \hline
   \textbf{Property} & \textbf{Symbol} & \multicolumn{4}{c}{\textbf{ETGs}}  & \textbf{Milky Way-like} \\
    \hline
    Stellar mass & $M_\star/M_\odot$ & $10^{10}$ & $10^{10.5}$ & $10^{11}$ & $10^{11.5}$ & $4.734 \times 10^{10}$ \\
    \hline
    Gas fraction & $M_{\rm gas}/M_\star$ & $7.86\times10^{-3}$ & $6.25\times10^{-3}$ & $1.43\times10^{-3}$ & $9.06\times10^{-4}$ & $1.09\times10^{-1}$ \\
    Star formation rate & $\dot M_{\star}/M_{\odot} \; {\rm yr^{-1}}$ & 1.02 & 2.83 & 0.34 & 3.32 & 4.08 \\
    Specific star formation rate & $\rm sSFR/yr^{-1}$ & $1.02\times10^{-10}$ & $9.03\times10^{-11}$ & $3.38\times10^{-12}$ & $1.06\times10^{-11}$ & $8.78\times10^{-11}$ \\
    Star formation efficiency & ${\rm SFE}/ {\rm yr^{-1}}$ & $1.29\times10^{-8}$ & $1.44\times10^{-8}$ & $2.36\times10^{-9}$ & $1.17\times10^{-8}$ & $7.55\times10^{-10}$ \\
  \hline
\end{tabular}
\caption{Global properties of each galaxy simulation at the snapshot used in this paper ($t=400$ Myr for the ETGs and $t=600$ Myr for the MW-like galaxy, both resulting in 300 Myr of analysis time), including stellar mass ($M_{\star}$), gas fraction ($M_{\rm gas}/M_\star$), total star formation rate ($\dot M_\star$), specific star formation rate (sSFR), and star formation efficiency (SFE; $\dot{M}_\star/M_{\rm gas}$). }\label{Tab::GalParams}
\end{centering}
\end{table*}

Introduced by \citet{Jeffreson2024b} (\citetalias{Jeffreson2024b}), the GalactISM suite of galaxy simulations consists of six chemodynamical isolated galaxies that collectively sample both the star-forming main sequence and quenched galaxy populations. Of the six original simulations, this work makes use of five: four early-type galaxies (ETGs) and one large spiral, or Milky Way-like system.

\begin{figure*}
\begin{centering}
    \includegraphics[width=\linewidth]{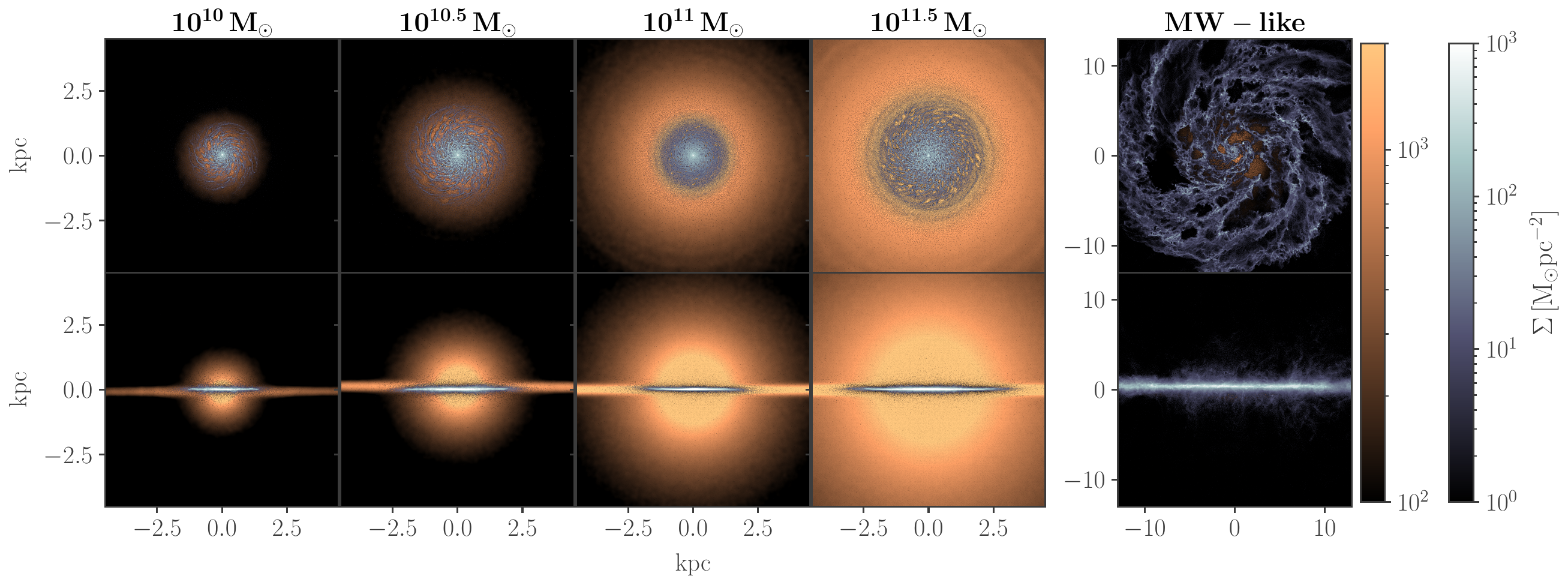}
    \caption{Stellar and gas surface density distributions across all galaxies in this paper (columns), with the ETGs increasing in mass from left to right, and the right-most column being the MW-like galaxy. First row depicts a face-on view of the galaxy while the second row is edge-on. The stellar distribution is represented by orange (first colorbar) and gas is grey/white (second colorbar). The MW-like galaxy has the thickest gas disk that extends beyond the stellar component. The suppressed ETG ($\rm 10^{11} \, M_{\odot}$) contains the smoothest and least disrupted gas disk.}
    \label{fig:gal_imshows}   
\end{centering}
\end{figure*}

\begin{table*} 
\begin{centering}
\vspace{0.5cm}
  \begin{tabular}{c c c c c c c}
  \hline
   \textbf{Property} &  \textbf{Symbol} & \multicolumn{4}{c}{\textbf{ETGs}}  & \textbf{Milky Way-like} \\
    \hline
    Stellar mass & $M_\star/M_{\odot}$ & $10^{10}$ & $10^{10.5}$ & $10^{11}$ & $10^{11.5}$ & $4.734 \times 10^{10}$ \\
    \hline
    \multicolumn{7}{c}{All overdensities} \\
    \hline
    Sobolev length & $\rm log \, (\ell$/pc) & 0.92 & 0.91 & 1.00 & 0.96 & 0.77 \\
    Gas mass & ${\rm log} \, (M_{\rm gas}/{\rm M_{\odot}})$ & 4.20 & 4.19 & 4.21 & 4.20 & 4.20 \\
    Freefall time & ${\rm log} \, (t_{\rm ff}/{\rm Myr})$ & 0.73 & 0.71 & 0.80 & 0.77 & 0.49 \\
    Crossing time & ${\rm log} \, (t_{\rm cross}/{\rm Myr})$ & -0.06 & -0.11 & 0.02 & -0.15 & -0.30 \\
    Virial parameter & ${\rm log} \, (\alpha_{\rm vir})$ & 1.62 & 1.67 & 1.68 & 1.88 & 1.63 \\
    Star formation rate & ${\rm log} \, (\dot M_{\star}/\rm M_{\odot} \; yr^{-1})$ & -11.79 & -12.54 & -9.34 & -11.90 & -9.51 \\
    Star formation efficiency & ${\rm log \, (SFE}/\rm Myr^{-1})$ & -8.74 & -9.45 & -6.33 & -8.84 & -6.45 \\
    Log total \# overdensities & - & 4.56 & 5.00 & 4.89 & 5.12 & 5.36 \\
    \hline
    \multicolumn{7}{c}{`Cloud-like' overdensities (SFE $>$ 0.01 $\rm Myr^{-1}$)} \\
    \hline
    Sobolev length & $\rm log \, (\ell$/pc) & 0.81 & 0.76 & 0.91 & 0.81 & 0.54 \\
    Gas mass & ${\rm log} \, (M_{\rm gas}/{\rm M_{\odot}})$ & 4.27 & 4.27 & 4.26 & 4.28 & 4.26 \\
    Freefall time & ${\rm log} \, (t_{\rm ff}/{\rm Myr})$ & 0.46 & 0.39 & 0.63 & 0.47 & 0.07 \\
    Crossing time & ${\rm log} \, (t_{\rm cross}/{\rm Myr})$ & 0.02 & -0.05 & 0.01 & -0.09 & -0.32 \\
    Virial parameter & ${\rm log} \, (\alpha_{\rm vir})$ & 1.03 & 1.02 & 1.39 & 1.22 & 1.03 \\
    Star formation rate & ${\rm log} \, (\dot M_{\star}/\rm M_{\odot} \; yr^{-1})$ & -4.19 & -4.14 & -4.49 & -4.18 & -4.35 \\
    Star formation efficiency & ${\rm log \, (SFE}/\rm Myr^{-1})$ & -1.28 & -1.23 & -1.60 & -1.26 & -1.42 \\
    Log total \# overdensities & - & 3.54 & 3.99 & 3.54 & 4.10 & 4.31 \\
    \hline
    \hline
\end{tabular}
\caption{Physical properties in logarithmic space of the individual overdensities in each galaxy simulation, including cloud radius ($\ell$), gas mass ($M_{\rm gas}$), freefall time ($t_{\rm ff}$), crossing time ($t_{\rm cross}$), virial parameter ($\alpha_{\rm vir}$), star formation rate ($\dot M_{\star}$), star formation efficiency (SFE), and the total number of respective overdensities. The first set of values corresponds to all identified overdensities in each respective galaxy (essentially the number of gas cells that cross the defined star formation threshold), while the second set of values represents overdensities that are more representative of realistic star-forming molecular clouds in the sense that they are more likely to be gravitationally bound, defined as having SFEs $> 0.01 \, \rm Myr^{-1}$. Values are medians over all overdensities in the corresponding galaxy, except for the total number. Distributions of all except SFE are visible in Figure~\ref{fig:overdensity_props}.}\label{Tab::Overdensity_Params}
\end{centering}
\end{table*}

Global properties of each galaxy at the snapshot used in this paper can be found in Table~\ref{Tab::GalParams}, while Figure~\ref{fig:gal_imshows} shows the stellar (orange) and gas (white) surface density distributions.

\subsection{Initial Conditions}\label{Sec::Sims_InitialConditions}

The process for selecting initial conditions for the GalactISM simulation suite is described in \citetalias{Jeffreson2024b}. We briefly recount this procedure here, though we refer readers to \citetalias{Jeffreson2024b} for further details.

All simulated galaxies draw their initial conditions from observations (see below), generated from {\sc MAKENEWDISK} \citep{Springel2005} or AGORA \citep{Kim2014}. 
No simulated galaxies in this suite (and subsequently this paper) include a circumgalactic medium (CGM; consistent with {\sc AGORA} initial conditions) or black holes. The properties of the dark matter halos, stellar disks and bulges, gas disks, and mass resolutions of particles can be found in Table 1 of \citetalias{Jeffreson2024b}. 

Though the ETGs and MW-like galaxy have different beginning and ending analysis times, as the beginning time is dependent on when the gas disk reaches equilibrium, all galaxies have an equivalent analysis timeframe of 300 Myr \citepalias{Jeffreson2024b}. The snapshots used in this paper (properties of which can be found in Table~\ref{Tab::GalParams}) represent the final snapshot of \citetalias{Jeffreson2024b}. We note that while this is only one instant in time, because of the idealized setup described below, we expect results to be the same even if another analysis snapshot is chosen.

\subsubsection{Early-Type Galaxies}

In order to simulate early-type galaxies that are comparable to observations, the parameters for the ETGs are sampled from the MASSIVE \citep{Ma2014} and $\rm ATLAS^{3D}$ \citep{Cappellari2011} surveys. In particular, ETG halo masses, stellar half-light radii, gas fraction, and the extent and surface density of the CO-luminous molecular gas disk are all sampled across logarithmic intervals in stellar mass ($M_\star=10^{10}, 10^{10.5}, 10^{11}, 10^{11.5} \,  M_{\odot}$). Similarly, all ETGs have a disk-to-bulge mass ratio of $M_{\rm \star,disk}/M_{\rm \star,bulge}=0.2$ for future comparisons with observations \citep{Utomo2015, Liu2021, Williams2023}. The dark matter halos of all ETGs are of Navarro-Frenk-White (NFW; \citealt{Navarro1997}) type with an exponential stellar and gas disk. The median gas cell mass is 859 $\rm M_{\odot}$.

The remaining properties, such as the ETG stellar bulge profiles, concentration parameters of dark matter halos, and spin parameter are set according to observational values from WISDOM and $\rm ATLAS^{3D}$ \citep{Cappellari2011, Ma2014}, described in Section 2.1 of \citetalias{Jeffreson2024b}.

\subsubsection{Milky Way-like Galaxy}

Intended to resemble a MW-like galaxy at $z=0$, initial physical properties include a stellar bulge that follows a \citet{Hernquist1990} profile with mass $3.4\times10^9 \, M_{\odot}$, disk mass of $4.3\times10^{10} \, M_{\odot}$, dark matter halo mass of about $10^{12} \, M_{\odot}$, and virial radius of 205 kpc. The halo concentration is $c=10$, and similar to the ETGs, has a spin parameter $\lambda=0.04$, NFW dark matter halo, exponential stellar and gas disk, and median gas cell mass of 859 $\rm M_{\odot}$.

\subsection{Physics Implementation}\label{Sec::Sims_Physics}

The initial conditions described above are evolved throughout the simulation runtime using {\sc AREPO}, a moving-mesh hydrodynamics code \citep{Springel2010}. The gaseous component is modeled using an unstructured moving Voronoi mesh constructed around discrete points, which move with the local fluid velocity. Gravitational accelerations for both Voronoi gas cells and particles (stellar and dark matter) are computed using a hybrid TreePM solver. 

\subsubsection{Chemistry}

The simulations adopt a nonequilibrium chemical network for hydrogen, carbon, and oxygen chemistry \citep{Nelson1997, Glover2007}, coupled to the atomic and molecular cooling functions of \citet{Glover2010} (see \citetalias{Jeffreson2024b} for details). Cooling channels include fine-structure emission from $\rm C^+$, O, and $\rm Si^+$; Ly$\alpha$ emission from atomic hydrogen; $\rm H_2$ line emission, gas-grain cooling; and electron recombination on grain surfaces and in reaction with polycyclic aromatic hydrocarbons (PAHs). At higher temperatures, the network also accounts for cooling from collisional processes, including $\rm H_2$ dissociation, bremsstrahlung, and ionization of atomic hydrogen. 

Heating is dominated by photoelectric emission from dust and PAHs, with some additional contribution from cosmic-ray ionization (rate of $2\times10^{-16} \; \rm s^{-1}$; \citealt{Indriolo2012}) and $\rm H_2$ photodissociation by the solar neighborhood-strength interstellar radiation field (ISRF; \citealt{Habing1968, Mathis1983}).
Dust abundances assume a solar dust-to-gas ratio, and temperatures follow the treatment of \citet{Glover2012}, Appendix A. 

\subsubsection{Star Formation}\label{Sec::Physics_SF}

Star formation within the gas cells is allowed based on density and temperature thresholds, where cells must have a temperature below 100 K and a density of at least $\rho_{\rm thresh}/m_{\rm H}\mu=100\;{\rm cm^{-3}}$, where $\mu$ is the mean mass per H atom. If these conditions are met, the SFR volume density follows the \citet{Padoan2017} model (see also \citealt{Gensior2020, Gensior2021}):

\begin{equation}
    \frac{d\rho_{\star,i}}{dt}=\frac{\epsilon_{\rm ff}\rho_i}{t_{{\rm ff},i}},
\end{equation}
where $t_{{\rm ff},i}=\sqrt{3\pi/(32G\rho_i)}$ is the local freefall timescale of the gas cell and $\rho_i$ is the cell's mass density. $\epsilon_{\rm ff}$ is defined as the efficiency of star formation per freefall time, and is dependent on the classical virial parameter such that
\begin{equation}\label{eqn::SFE_per_tff}
    \epsilon_{\rm ff}=0.4{\rm exp}(-1.6\alpha_{\rm vir}^{0.5}).
\end{equation}

The virial parameter $\alpha_{\rm vir}$ is calculated on a scale $\ell$ as
\begin{equation}\label{eqn::virial_param}
    \alpha_{\rm vir} = 1.35\left(\frac{2t_{\rm ff}\sigma|\nabla \rho|}{\rho^{3/2}}\right)^2,
\end{equation}
where $t_{\rm ff}$ is now the freefall time over the entire overdensity within radius $\ell$ (as opposed to the individual gas cell), and $\sigma$ is the 3-D velocity dispersion, while $\rho$ and $|\nabla \rho|$ come from a variant of the Sobolev approximation \citep{Sobolev1960} used to determine the overdensity size as $\ell = \rho/|\nabla\rho|$. The Sobolev length, $\ell$, therefore estimates the distance over which the density changes considerably, meaning that overdensities are individually measured relative to the local environment as opposed to some pre-selected uniform extent. A thorough explanation of the Sobolev length calculation can be found in Section 2.2 of \citet{Gensior2020}, but we highlight here that this length-scale is better suited to the dynamic and multiphase ISM, and is largely invariant with resolution (Appendix A of \citealt{Gensior2020}). We briefly discuss how our results are affected by a fixed uniform scale in Section~\ref{Sec::Discussion_Caveats}. 

\subsubsection{Feedback}

To calculate the momentum and energy released by supernovae and pre-supernova HII region feedback, stellar populations are stochastically sampled from a \citet{Chabrier2003} initial stellar mass function (IMF) using the Stochastically Lighting Up Galaxies (SLUG) stellar population synthesis model \citep{DaSilva2014, Krumholz2015}. From this, the number of supernovae, ejected mass, and photoionizing luminosity of each star are computed, assuming that each supernova releases an energy of $10^{51}$ ergs, and the terminal supernova momentum is explicitly calculated from Equation 17 of \citet{Gentry2017}. The kinetic energy and residual thermal energy are injected into gas cells surrounding each star particle.

HII region photoionizing luminosities are translated into momentum input using the prescriptions of \citet{Jeffreson2021}, which incorporate both radiation pressure and the ``rocket effect", which is the ejection of warm ionized gas from cold molecular clouds \citep{Matzner2002, Krumholz2009}. Any gas within the Stromgren radii is fully ionized and heated to 7000 K. 

\section{Modified Virial Theorem}\label{Sec::VirialTheorem}

To investigate the driving forces behind star formation (or lack thereof) in the simulated galaxies, we identify overdensities (`clouds') within the galaxies and compute the physical forces acting on and within these regions from a modified version of the virial theorem (Modified Virial Theorem; MVT). 

One of the most general forms of the virial theorem (for some volume) can be stated as 
\begin{equation}\label{eqn::MVT1}
    \frac{\ddot I}{2} = 2E_k + \int_V \left(\vec{a}(\vec{d})\cdot\vec{d}\right)dm,
\end{equation}
where, in this particular case, $\vec{d}$ is the displacement vector from the center of mass (COM), $\ddot{I}$ is the moment of inertia of the overdensity, $E_k$ is the total kinetic energy of the gas, and the integral is the volume integral over the overdensity. Within the integral, $\vec{a}=\ddot{\vec{d}}$, or the acceleration relative to the center of mass (COM) of the overdensity. This form of the virial theorem captures the internal forces between components within the system. 

However, motivated by \citet{Meidt2018} and \citet{Liu2021}, this general virial theorem can be modified to include work done by external forces (see also \citealt{Binney2008}).
In this paper, we focus on possible external forces acting through the gravitational potential, and split them by direction, including a term acting within the midplane ($U_{\rm R,ext}$) and a term acting perpendicular to the midplane ($U_{\rm z,ext}$). Therefore, by including these external forces in addition to the kinetic energy $E_k$ and self-gravity $U_{\rm sg}$ of the classical virial theorem, we get a modified virial theorem:
\begin{equation}
    \frac{\ddot I}{2} = 2E_k + U_{\rm sg} + U_{\rm R,ext} + U_{\rm z,ext},
\end{equation}

This essentially computes the net projection of the forces onto the vector that points toward the center of mass of the overdensity, $\vec{d}$, and the sum of these projections tells us whether gas is being pushed towards the center (compression) or away (support against collapse/expansion). As stated by \citet{Liu2021}, if the time-averaged value of $\ddot{I}$ is equal to zero, the overdensity is in equilibrium, neither expanding nor collapsing. 

The conditions required for collapse, using our modified virial theorem, can then be written as:
\begin{equation}\label{eqn::MVT2}
    0 > 2E_{\rm k} + U_{\rm sg}+U_{\rm R,ext}+U_{\rm z,ext}
\end{equation}
where each $U$ term is the energy contribution from self-gravity ($U_{\rm sg}$), external gravity in the midplane ($U_{\rm R, ext}$), and external gravity perpendicular to the midplane ($U_{\rm z, ext}$), respectively. Then, in the form of Equation~\ref{eqn::MVT1},
\begin{equation}\label{eqn::MVT3}
\begin{split}
    0 > 2E_{\rm k} + \int_V \left(\vec{a}_{\rm sg}(\vec{d})\cdot \vec{d}\right) dm & + \int_V \left(\vec{a}_{\rm R,ext}(\vec{d})\cdot \vec{d}\right) dm \\ & + \int_V \left(\vec{a}_{\rm z,ext}(\vec{d})\cdot \vec{d}\right) dm.
\end{split}
\end{equation}

To calculate the terms on the right hand side, one-by-one, for a system of discrete particles (gas cells within a specific overdensity), we can write them as sums over the gas cell properties, such as masses $m_i$ or positions $(x_i, y_i, z_i)$ with respect to the galactic center. 

For the total kinetic energy $E_k$, this results in:

\begin{equation}\label{eqn::MVT4_kineticenergy}
    E_{\rm k} = \frac{1}{2}\sum_i w_im_i  (\vec{v}_i - \langle  \vec{v}_i \rangle)^2 
\end{equation} 
where $w_i$ is the weight from a cubic spline kernel (see Section 7 of \citealt{Monaghan1992}), $m_i$ is the mass, $v_i$ is the velocity, and $\langle \vec{v}_i\rangle$ is the kernel-weighted mean of all overdensity gas velocities. 

Of this total kinetic energy $E_{\rm k}$, the amount contributed by the rotational kinetic energy $E_{\rm k, rot}$ (such that the rotational kinetic energy is always some fraction of the total kinetic energy, i.e., $E_{\rm k,rot}/E_{\rm k} < 1$) is:

\begin{equation}\label{eqn::MVT5_rotational_kineticenergy}
    E_{\rm k,rot} = \frac{1}{2}\sum_i w_im_i v_{\phi,i}^2
\end{equation} 
where $v_{\phi,i}$ is the tangential velocity relative to the overdensity's center of mass. 

The self-gravity contribution is

\begin{equation}\label{eqn::MVT6_selfgravity}
    U_{\rm sg} = \sum_i w_i m_i \sum_{j\neq i} -\frac{Gm_j}{|\vec{r}_{ij}|^3} \vec{r}_{ij} \cdot \vec{d}_i,
\end{equation}
where $\vec{r}_{ij}=(x_{ij}, y_{ij}, z_{ij})$ is the vector from gas cell $i$ to cell $j$, and $\vec{d}_i=(d_{x,i}, d_{y,i}, d_{z,i})$ the vector from the center of mass of the overdensity to gas cell $i$.

Next, the external gravitational forces $U_{\rm z,ext}$ and $U_{\rm R,ext}$ are derived from the gravitational potential (calculated as described in Appendix B of \citetalias{Jeffreson2024b}). The simpler vertical component is:

\begin{equation}\label{eqn::MVT7_zext}
    U_{\rm z,ext} = \sum_i w_i m_i \left(-\frac{\partial \Phi_{\rm ext}}{\partial z}|_i d_{z,i}\right)
\end{equation}
where $\frac{\partial \Phi_{\rm ext}}{\partial z}|_i$ is the vertical gradient of the gravitational potential $\Phi_{\rm ext}$ at the position of gas cell $i$, excluding the contribution of the overdensity itself. 

The in-plane external component can be split into an axisymmetric and non-axisymmetric part:

\begin{equation}\label{eqn::MVT8_Rextax}
\begin{split}
 U_{\rm R,ext,ax} = \sum_i w_i m_i & \left[ \left. \frac{-\partial\Phi_{\rm ext}}{\partial x} \right|_i d_{x,i} \, - \, \left. \frac{\partial \Phi_{\rm ext}}{\partial y}\right|_i d_{y,i}\right. \\ & + \left. \, \Omega_0^2 \left( x_i d_{x,i} + y_{i}d_{y,i} \right) \right]
\end{split}
\end{equation}
where the partial derivatives make up the `tidal' force ($U_{\rm R,ext,tidal}$) and the last term is the centrifugal force ($U_{\rm centrifugal}$). We again note that $(x_i, y_i)$ are the in-plane positions of the gas cell with respect to the \emph{galactic} center, while ($d_{x,i}$, $d_{y,i}$) are positions relative to the \emph{overdensity's} center of mass. $\Omega_0$ is the angular velocity of the overdensity's circular orbit at the specified radius $\left(v_c(R)/R\right)$. The gradients in the tidal term describe how variations in the galactic field across the overdensity act to stretch it or compress it in the plane. The centrifugal term represents the outward force due to the overdensity's orbital rotation around the galactic center.

Finally, the non-axisymmetric contribution is

\begin{equation}\label{eqn::MVT9_Rextnax}
    U_{\rm R,ext,nax} = \sum_i w_i m_i \left[ -2\Omega_0 (\dot{d}_{x,i} d_{y,i} - \dot{d}_{y,i} d_{x,i}) \right].
\end{equation}

This term is effectively the contribution from the Coriolis force, and we hereafter use $U_{\rm R,ext,nax}$ and $U_{\rm Coriolis}$ interchangeably. This term quantifies how the overdensity's rotation around the galaxy couples to the motions of the gas within it (relative to the cloud COM), similar to shear, contributing to the overall rotational energy budget.

In the coming sections, we will compare the relative magnitudes of the energy terms ($U_{\rm sg}$, $U_{\rm R,ext}$, etc.) for the overdensities of the different galaxies in our simulation suite, focusing in particular on any differences displayed by the one ETG in our sample that has a suppressed star formation rate.

\section{Results}\label{Sec::Results}

To determine what is acting to support overdensities against collapse, and how the suppressed ETG varies from its GalactISM counterparts, we identify all overdensities within the galaxies and employ the MVT as detailed in Section~\ref{Sec::VirialTheorem}. `Overdensities' are hereby defined as the region around gas cells that meet the requirements for star formation (see Section~\ref{Sec::Physics_SF}). As was done during the simulation run, the size of an overdensity is determined by the Sobolev length $\ell=\rho/|\nabla\rho|$. This length scale is not constant and varies across all galaxies.

As in \citetalias{Jeffreson2024b}, we exclude any gas cells with distances less than $R_{\rm min}=50$ pc from the galactic center, and impose a maximum distance $R_{\rm max}=1.5$ kpc for the ETGs, and $R_{\rm max}=13$ kpc for the MW-like galaxy. These maximum radii ensure that the majority of the gas disks are included for each galaxy.

\subsection{Characterization of Overdensities}\label{Sec::overdensity_characterization}

\begin{figure*}
\begin{centering}
    \includegraphics[width=\linewidth]{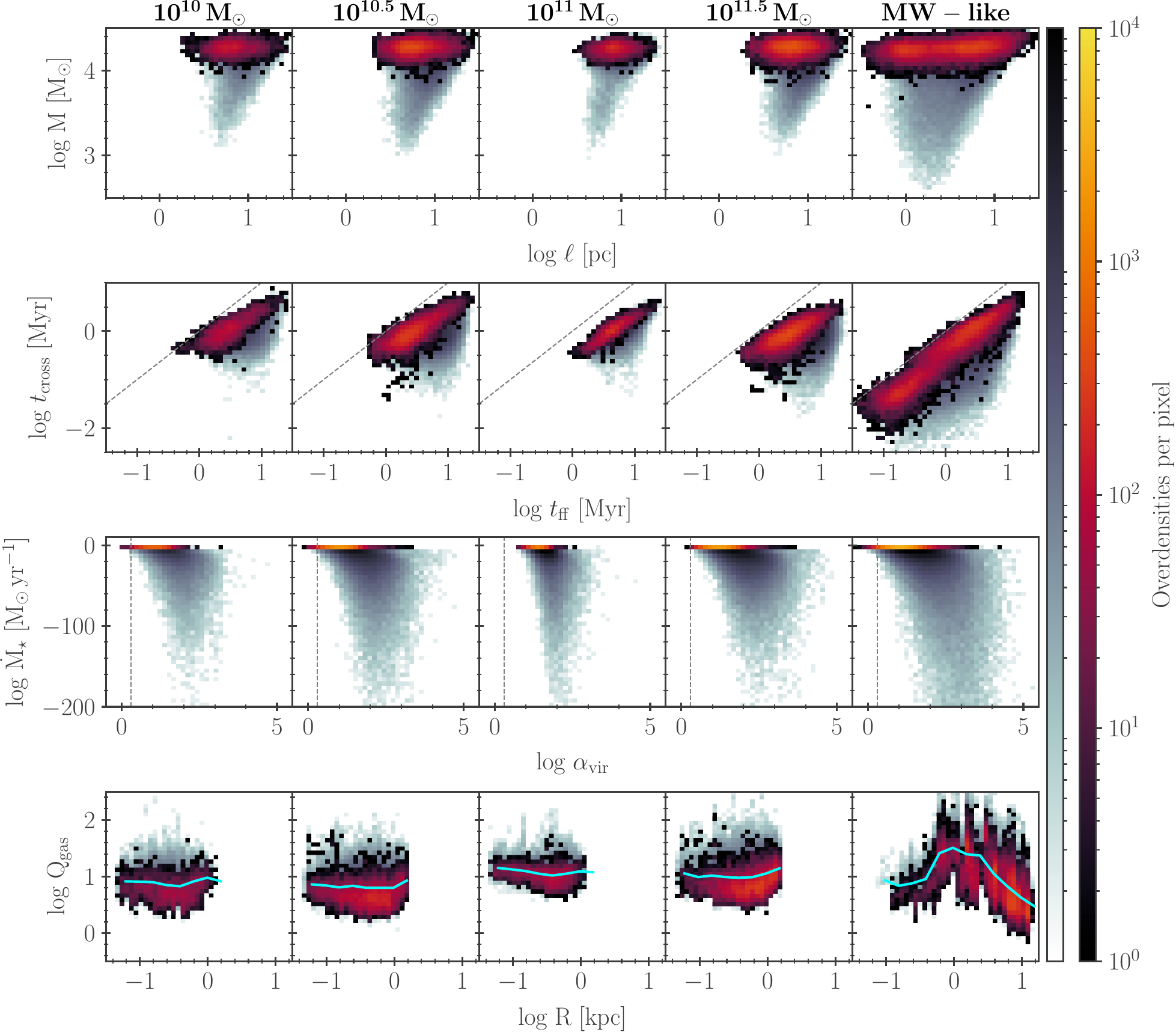}
    \caption{2-D histograms of overdensity properties across all galaxies in this paper (columns), with the ETGs increasing in mass from left to right, and the right-most column being the MW-like galaxy. First row depicts the Sobolev length $\ell$ (cloud radius) versus total mass enclosed $M$, second row is the cloud freefall time $t_{\rm ff}$ versus crossing time $t_{\rm cross}$, third row is the classical virial parameter $\alpha_{\rm vir}$ (from Equation~\ref{eqn::virial_param}) versus SFR $\dot M_\star$, and last row is the classical Toomre-$Q$ parameter $Q_{\rm gas}$ and radius of the overdensity from the galactic center $R$. The dashed grey line in the second row represents where $t_{\rm cross}=t_{\rm ff}$, while in the third row it represents where $\alpha_{\rm vir}=2$. Cyan line in the bottom row represents the median value in the gas across the galaxy, similar to Figure 7 of \citetalias{Jeffreson2024b}. The second colorbar denotes data where the SFE is at least $0.01 \;\rm Myr^{-1}$, ideally being more representative of realistic gravitationally-bound clouds. The suppressed ETG shows larger Sobolev lengths $\ell$. Both the suppressed ETG ($\rm 10^{11} \, M_{\odot}$) and highest-mass ETG ($\rm 10^{11.5} \, M_{\odot}$) show no overdensities with crossing times larger than the freefall time. The suppressed ETG is the only galaxy to have no overdensities with $\alpha_{\rm vir}\leq 2$. }
    \label{fig:overdensity_props}   
\end{centering}
\end{figure*}

We first characterize the properties of each galaxy's overdensities. Table~\ref{Tab::Overdensity_Params} contains the the median values of each galaxy's overdensity physical parameters, including the size ($\ell$), gas mass contained within ($M_{\rm gas}$), freefall time ($t_{\rm ff}$), crossing time ($t_{\rm cross}$), virial parameter ($\alpha_{\rm vir}$), the star formation rate of the central star-forming gas cell ($\dot M_\star$), and the star formation efficiency (SFE) of this central cell. The distributions of all overdensities within each galaxy, for the parameters just described (excluding SFE), are shown in Figure~\ref{fig:overdensity_props}.

First, we note that while identifying overdensities with such high $\alpha_{\rm vir}$ that are not gravitationally bound may seem counterintuitive to the definition of a star-forming region, we remind the reader that we define overdensities as the region around every star-forming cell (see Section~\ref{Sec::Physics_SF}), as opposed to objects more reminiscent of giant molecular clouds (GMCs). Cells are deemed `star-forming' if they exceed a certain density threshold, discussed in Section~\ref{Sec::Physics_SF}, rather than a certain virial parameter. The SFE is then determined via the virial parameter (Equation~\ref{eqn::SFE_per_tff}). As a result, for overdensities with significant (cloud-like) SFEs, the virial parameters will also be cloud-like, but for general overdensities, there is no guarantee that the velocity dispersion is conducive to collapse. 

Across the ETGs, the median overdensity size from the Sobolev length $\ell$ is similar, being on the order of $\ell \approx 10$ pc. The suppressed ETG ($\rm 10^{11} \, M_{\odot}$), however, maintains the highest median `cloud' size of about 12 pc. The MW-like galaxy has $\ell$ of nearly half that, with a median of 5.91 pc. From the top row of Figure~\ref{fig:overdensity_props}, it becomes apparent that the suppressed ETG entirely lacks star-forming overdensities with Sobolev lengths below $\sim$6.5 pc. In addition, while the median overdensity masses are similar across all galaxies ($\approx \rm 1.6\times10^6 \, M_{\odot}$), the suppressed ETG lacks overdensities of lower masses compared to the rest of the simulation suite. 

These differences are also pronounced when comparing the freefall time ($t_{\rm ff}=\sqrt{3\pi/(32G\rho)}$) and turbulent crossing time ($t_{\rm cross}=R_{\rm cl}/2\sigma$; \citealt{Bertoldi1992, Gensior2020}). In Figure~\ref{fig:overdensity_props}, all galaxies predominantly show significantly larger $t_{\rm ff}$ than $t_{\rm cross}$, with some galaxies appearing to have no data where $t_{\rm ff}\leq t_{\rm cross}$, most pronounced in the suppressed ETG. As mentioned above, due to our definition of overdensities and star formation prescriptions, $t_{\rm ff}\gg t_{\rm cross}$ is to be expected, especially in the ETGs.

This indicates that most overdensities we identify are likely not gravitationally bound, which is supported by  the values for the virial parameter (Table~\ref{Tab::Overdensity_Params}), which range from a median of about 35 for the $\rm 10^{10}M_{\odot}$ ETG up to about 72 for the $\rm 10^{11.5}M_{\odot}$ ETG. Of all the galaxy simulations used in this paper, the suppressed ETG is the only one to have no overdensities with $\rm \alpha_{\rm vir}\leq 2$.

\subsection{Overdensity Virial Properties}

With an understanding of the basic physical properties of the overdensities in these galaxies, we now use the MVT (see Section~\ref{Sec::VirialTheorem}) to conduct an analysis into the forces acting for or against cloud collapse. For simplicity, we present seperately the energy terms from kinetic energy and self-gravity (Section~\ref{Sec::KE_and_SG_forces}), the components in the midplane, (Section~\ref{Sec::midplane_forces}), and the component perpendicular to the midplane (Section~\ref{Sec::vertical_forces}).

Throughout the rest of this paper, we include results for all overdensity distributions (as in Figure~\ref{fig:overdensity_props}) and for a second set of distributions, limited to overdensities with a star formation efficiency $\dot M_\star/M_{\rm gas} > 0.01 \, {\rm Myr^{-1}}$. This second set of distributions limited by SFE is more likely to be representative of gravitationally-bound systems according to the traditional version of the virial parameter $\alpha_{\rm vir}$ \citep{Krumholz2005, Volschow2017, Grudic2019, KimJ2021}, which determines the SFE directly during the simulation via the \citet{Padoan2017} model.

This set of overdensities is of particular interest for two reasons. First, overdensities with SFE $>0.01 \rm \, Myr^{-1}$ account for $>$97\% of the star formation in most simulated galaxies ($>$85\% for the suppressed ETG), and therefore the physics impacting these overdensities are consequently the physics impacting star formation in each galaxy. Second, overdensities with high $\alpha_{\rm vir}$ are already dominated by the turbulent kinetic energy, and thus the additional expansive effect of external gravitational forces is far less relevant.

From the last row of Figure~\ref{fig:overdensity_props}, we reiterate the fact that the gaseous Toomre-$Q$ parameter is similarly high in all ETGs ($Q_{\rm gas} \approx 10$), and is not significantly elevated in the suppressed galaxy.

\subsubsection{Kinetic Energy and Self-Gravity Components}\label{Sec::KE_and_SG_forces}

\begin{figure*}
\begin{centering}
    \includegraphics[width=\linewidth]{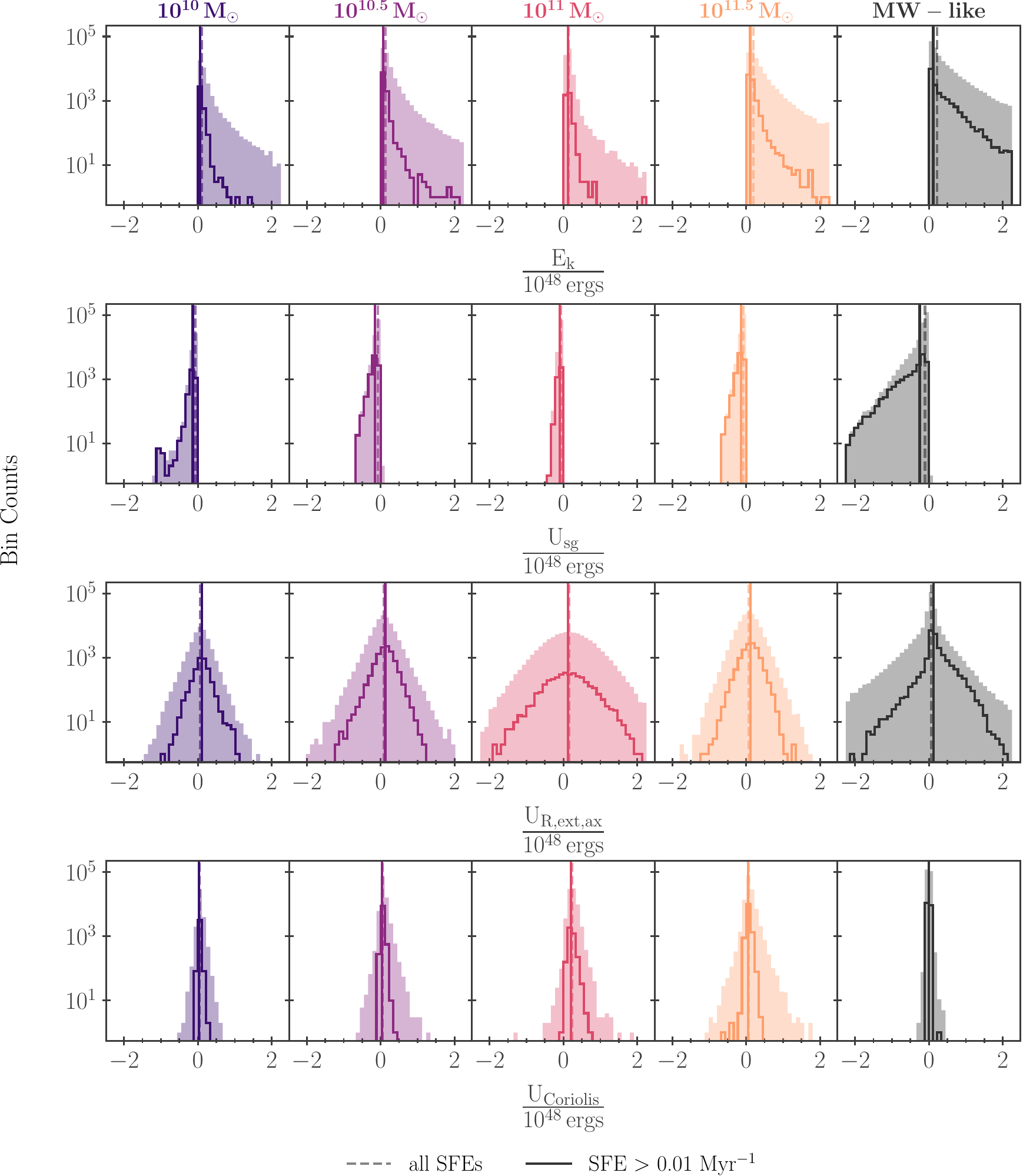}
    \caption{Distributions of total kinetic energy (first row), self-gravitational energy (second row), axisymmetric external midplane energy (third row), and energy from the Coriolis force (fourth row), from Equations~\ref{eqn::MVT4_kineticenergy}, \ref{eqn::MVT6_selfgravity}, \ref{eqn::MVT8_Rextax}, and \ref{eqn::MVT9_Rextnax}, respectively. Lighter filled histograms represent all overdensities, and the dashed vertical line is the median. Darker unfilled histograms (solid lines) show the distribution for all overdensities with a star formation efficiency (SFE) greater than 0.01 $\rm Myr^{-1}$, which are more likely to represent gravitationally-bound systems, while the solid vertical line is the median for such systems. }
    \label{fig:all_force_hist}   
\end{centering}
\end{figure*}

The original terms of the classic virial theorem, kinetic energy and self-gravity are the two primary competitors in a simplified scenario of cloud collapse and eventual star formation, and form the definition of the virial parameter \citep{Padoan2017}. Our distributions of the total kinetic energy (from Equation~\ref{eqn::MVT4_kineticenergy}) and energy from self-gravity (Equation~\ref{eqn::MVT6_selfgravity}) can be found in the first two rows of Figure~\ref{fig:all_force_hist}.

From the top row of Figure~\ref{fig:all_force_hist}, the medians of the total kinetic energy distributions in the ETGs, regardless of SFE, appear to increase with stellar mass, albeit not evenly. Of all simulations, the MW-like galaxy shows the largest values of kinetic energy. Though we do not show it here, the rotational kinetic energy in all galaxies shows medians of about a third of the total kinetic energy.

The only purely compressive (negative) term in this paper, the energy contributions from self-gravity are indicated in the second row of Figure~\ref{fig:all_force_hist}. The self-gravities appear to show a clearer difference not only between the galaxies themselves, but also between overdensities of varying SFE. Furthermore, the magnitudes of self-gravity are significantly smaller than the kinetic energies.

As might be expected, overdensities with higher SFEs show larger self-gravities. However, unlike the kinetic energy, medians of self-gravity do not scale with ETG stellar mass: the $10^{11}\,{\rm M_{\odot}}$ ETG has the weakest (and narrowest) self-gravities, while the $10^{10.5}\,{\rm M_{\odot}}$ ETG has the strongest with values narrowly larger than the highest-mass ETG. The MW-like galaxy boasts the largest self-gravities with medians almost twice that of the $10^{10.5}\,{\rm M_{\odot}}$ ETG, and has the most deviation between sample medians. Again, the suppressed ETG shows the least variation when considering a limit on SFE, but this variation in self-gravity medians ($\approx 2.7\times10^{46}$ ergs) is still twice that of the kinetic energy.

We note that the lack of stronger disparity between the suppressed ETG's distributions of kinetic energy and self-gravity when limiting by SFE (shaded/unshaded histograms and solid/dashed vertical lines of Figures~\ref{fig:all_force_hist}) may be representative of the fact that the galaxy contains so few gravitationally-bound systems, as mentioned previously in Section~\ref{Sec::overdensity_characterization}.

\subsubsection{Midplane Energy Components}\label{Sec::midplane_forces}

As part of the addition to the classical virial theorem, we include consideration of energy contributions in the midplane. These terms consist of axisymmetric contributions from the external gravity, $U_{\rm R,ext,ax}$ defined by Equation~\ref{eqn::MVT8_Rextax}, and the non-axisymmetric energy from the Coriolis force, $U_{\rm Coriolis}$ ($U_{\rm R,ext,nax}$ in Equation~\ref{eqn::MVT9_Rextnax}).

As with previous energy components, the distributions for $U_{\rm R,ext,ax}$ are visible in the third row of Figure~\ref{fig:all_force_hist}. Each galaxy shows an approximately normal distribution with medians slightly larger than $U_{\rm R,ext,ax}=0$. All galaxies show little variation when considering ${\rm SFE}>0.01\,{\rm Myr^{-1}}$; the most notable difference is that the suppressed ETG has the widest range of values. 

However, we note that there is some nuance in the $U_{\rm R,ext,ax}$ term, as it would be expected to be zero at the center of mass in a circular orbit, which we discuss further in Appendix~\ref{appendix::Rextaxi_components}. For this reason, as in \citet{Meidt2018}, we do not consider this energy term in the rest of this paper.

The final midplane energy term, the non-axisymmetric contribution from the Coriolis force, is arguably the most important presented in this paper. The last row of Figure~\ref{fig:all_force_hist} shows that, like the axisymmetric term in the third row, these distributions have little variation in medians with SFE. However, each histogram has a slight positive skew. This effect is less pronounced in the Milky Way-like galaxy, which maintains a median of almost exactly zero, but is more pronounced in the ETGs. Each ETG's median is again slightly offset from zero, with the exception of the suppressed ETG, which has a median energy contribution from the Coriolis force 4-6 times larger than that of the other ETGs.

Due to this notable difference, we continue our analysis of the Coriolis contribution beginning in Section~\ref{Sec::overdensities_and_coriolis}. 

\subsubsection{Vertical Energy Component}\label{Sec::vertical_forces}

The last modification to the original virial theorem takes the form of $U_{\rm z,ext}$, the energy contribution from the external gravitational potentials in the $z$-direction (perpendicular to the midplane), as defined by Equation~\ref{eqn::MVT7_zext}. These distributions are visible in Appendix~\ref{appendix::vertical_components}.

Because the median value of $U_{\rm z,ext}$ in every galaxy is approximately zero, representing the lowest energy contribution in the MVT, we briefly elaborate on this energy in Appendix~\ref{appendix::vertical_components}, but ultimately do not further consider it in our analysis.

\subsection{Overdensity Properties and the Coriolis Force}\label{Sec::overdensities_and_coriolis}

From the energy terms in the galaxy overdensities just presented, it is clear that there are two that stand out in regards to the suppressed ETG: $U_{\rm sg}$ (Equation~\ref{eqn::MVT6_selfgravity}) and $U_{\rm Coriolis}$ (Equation~\ref{eqn::MVT9_Rextnax}). To understand how these contribute to star formation, the relevant physical terms setting the simulation SFR must be analyzed.

From Equation~\ref{eqn::SFE_per_tff}, the star formation rate is ultimately dependent on the virial parameter $\alpha_{\rm vir}$, which, from \citet{Padoan2017}, is intended to represent an approximate ratio of kinetic energy to self-gravity. A gravitationally-bound, collapsing overdensity would therefore be expected to have lower $\alpha_{\rm vir}$, typically $\leq 2$. Therefore, while the self-gravity is then naturally connected to $\alpha_{\rm vir}$, if the Coriolis force has a true influence on the star formation rate, we would expect to see some relationship with $\alpha_{\rm vir}$. 

\begin{figure*}
\begin{centering}
    \includegraphics[width=\linewidth]{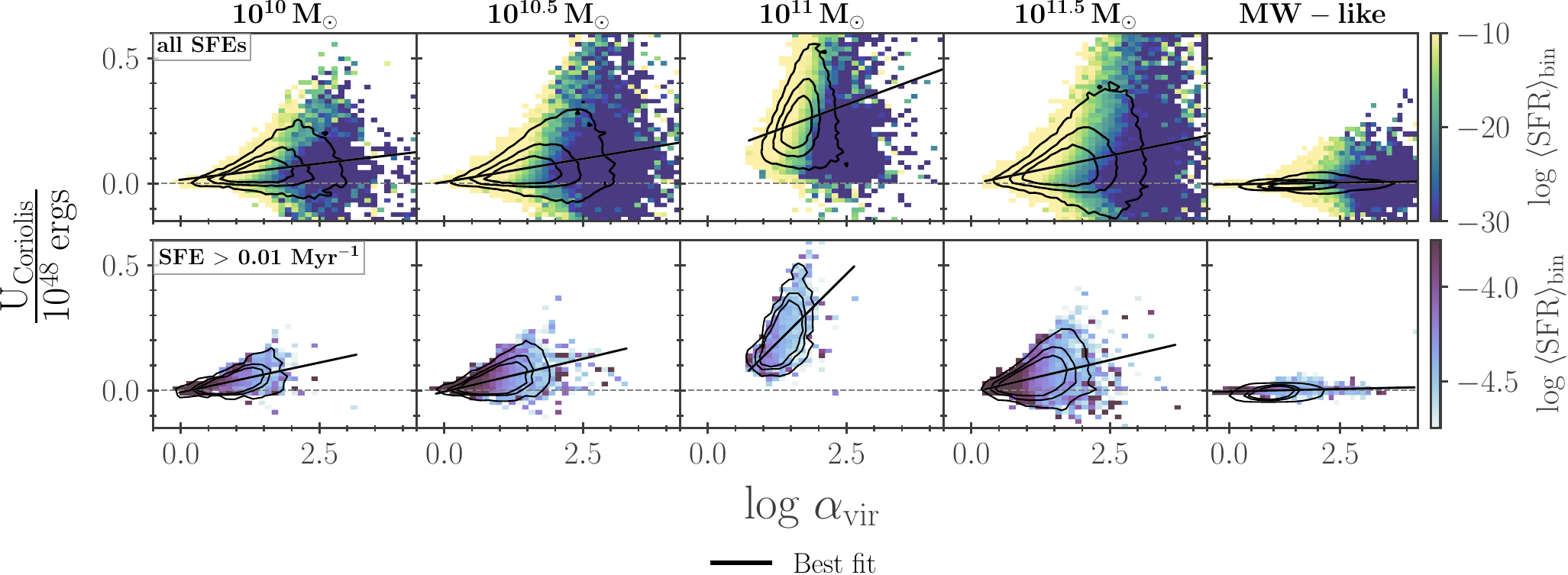}
    \caption{2-D histograms of the classical virial parameter from Equation~\ref{eqn::virial_param}, $\alpha_{\rm vir}$, versus the energy term from the Coriolis force, colored by the mean SFR in each bin. Unfilled contours show the 50\%, 75\%, and 95\% data inclusion regions, while the solid line is the least squares regression. Dashed grey line is where the Coriolis energy term is 0. The top row is all overdensities, while the bottom row is limited to overdensities with SFE $> 0.01 \; {\rm Myr^{-1}}$. The suppressed ETG ($\rm 10^{11}M_{\odot}$) is the only galaxy with an energy contribution from the Coriolis force that is almost entirely positive, and maintains a much steeper slope than the other ETGs. The MW-like galaxy shows little variation in the Coriolis force with $\alpha_{\rm vir}$, remaining extremely flat. }
    \label{fig:alpha_vs_coriolis}   
\end{centering}
\end{figure*}

The actual relationships between the classical virial parameter, $\alpha_{\rm vir}$ from Equation~\ref{eqn::virial_param}, and the Coriolis energy contribution for each galaxy are visible in Figure~\ref{fig:alpha_vs_coriolis}. Pixels are colored by the log mean SFR within that respective bin, and unfilled contours show the inclusion of 50\%, 75\%, and 95\% of the data distribution, respectively. The light grey dashed line is indicative of $U_{\rm Coriolis}=0$, while the solid black line is the least squares regression (line of best linear fit).

As expected from the star formation prescriptions in the GalactISM suite, Figure~\ref{fig:alpha_vs_coriolis} shows an increase in the mean SFR of each bin (lighter colors in top row, darker colors in bottom row) as the virial parameter decreases. However, whether the samples are limited by star formation efficiency or not (top versus bottom rows), relationships between galaxy types remain the same. 

The MW-like galaxy shows a much flatter relationship between $U_{\rm Coriolis}$ and $\alpha_{\rm vir}$ than any of the ETGs, as most of the distribution is centered around $U_{\rm Coriolis}=0$, as demonstrated in Figure~\ref{fig:all_force_hist}. On the other hand, all ETGs, with the exception of the $10^{11}\, M_{\odot}$ suppressed galaxy, show nearly identical distributions in $\alpha_{\rm vir}$ and $U_{\rm Coriolis}$ with a positive slope. This suppressed galaxy deviates by showing a much steeper positive slope, indicating a sharper rise in the Coriolis energy contribution with increasing virial parameter. 

Next, to investigate how the Coriolis energy contribution affects our MVT presented in Section~\ref{Sec::VirialTheorem}, we now compare the sum of the Coriolis force's energy with the other relevant terms. As mentioned earlier, if the vertical external energies are neglected due to low magnitudes and symmetry, and the axisymmetric midplane forces are neglected due to the expectation that they should be negligible, the only terms left in the MVT are the kinetic energy $E_k$, self-gravity $U_{\rm sg}$, and Coriolis energy, $U_{\rm Coriolis}$. While it has been established that the suppressed ETG lacks self-gravities compared to the other ETGs, and exhibits larger Coriolis energies, we now investigate these summed energy terms. 

\begin{figure*}
\begin{centering}
    \includegraphics[width=\linewidth]{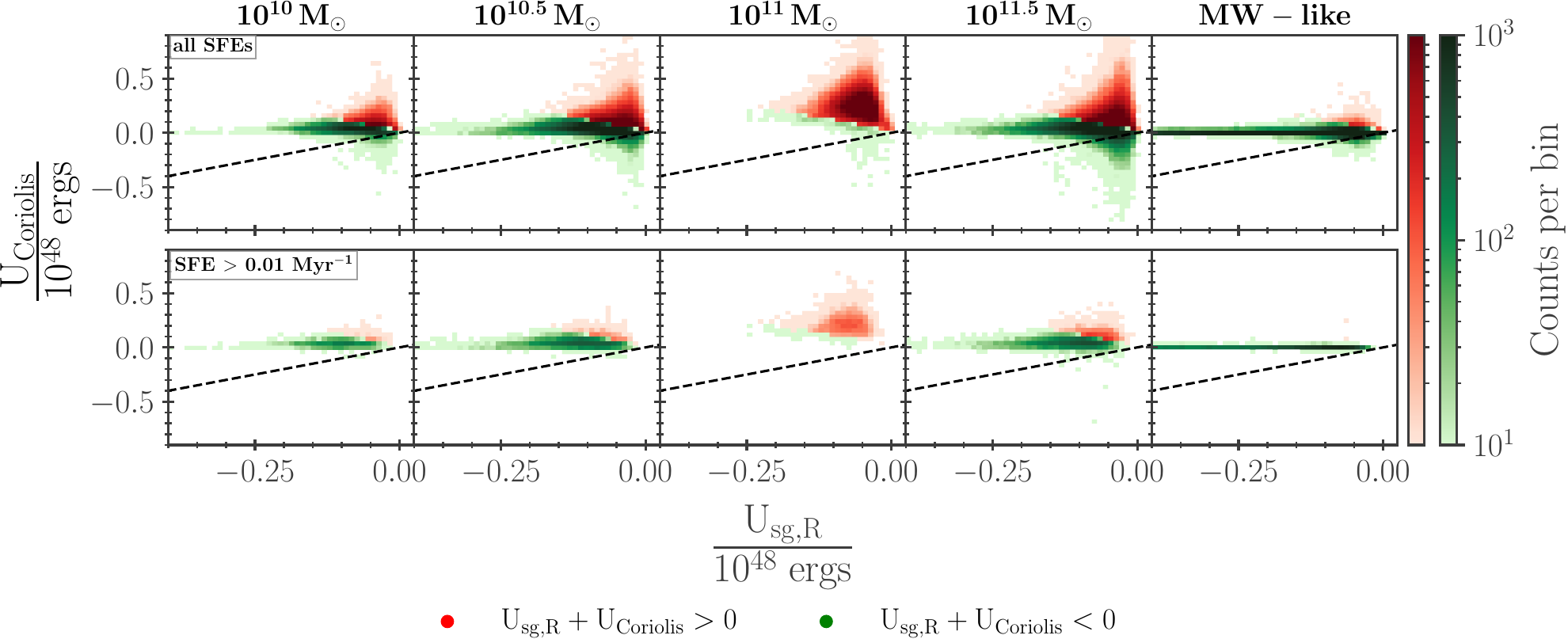}
    \caption{2-D histograms of the self-gravity in the midplane direction versus the Coriolis energy term. Pixels are colored based on the sum of the force terms (x- and y-axis), where reds represent a positive sum (support against collapse) and greens are a negative sum (collapsing overdensity). The dashed black line represents where the two axes are equal. As in Figure~\ref{fig:axi_anticorrelation}, the top row is all overdensities, while the bottom row is limited to overdensities with SFE $> 0.01 \; {\rm Myr^{-1}}$. The suppressed ETG ($\rm 10^{11} M_{\odot}$) shows the least amount of collapsing overdensities (green).}
    \label{fig:SG_vs_coriolis}   
\end{centering}
\end{figure*}

Figure~\ref{fig:SG_vs_coriolis} compares the self-gravity in the radial (midplane) direction with the energy from the Coriolis force. Pixels are colored depending on their sum: if the net energy term of $U_{\rm sg,R}+U_{\rm Coriolis}$ is positive, indicating support against collapse, the pixel is red. Otherwise, if the energy sum is negative, potentially indicating a collapsing overdensity, the pixel is colored green. In both cases, the color shade represents the number of overdensities within that particular bin; the only difference between colors is the sign of the net energy term. The black dashed line differentiates between these two samples. 

We note that only self-gravity from the radial direction is considered in this figure due to negligible contributions in the $z$-direction, and for the most comparable analysis between midplane forces. Using the total self-gravity term (as presented in Figure~\ref{fig:all_force_hist}) would yield similar results.

From Figure~\ref{fig:SG_vs_coriolis}, it again becomes apparent that there are several distinct patterns depending on the galaxy. The MW-like galaxy has a strong self-gravity component, leading to a majority of overdensities being net compressive (green). In the ETGs, the amount of both compressive and expansive overdensities appears to increase with stellar mass, with the exception of the $\rm 10^{11}\,M_{\odot}$ ETG. In this case, the suppressed ETG maintains few compressive overdensities, even at higher SFEs. For the rest of the ETGs, the higher SFE overdensities appear to be concentrated around much lower $U_{\rm Coriolis}$.

\begin{figure*}
\begin{centering}
    \includegraphics[width=\linewidth]{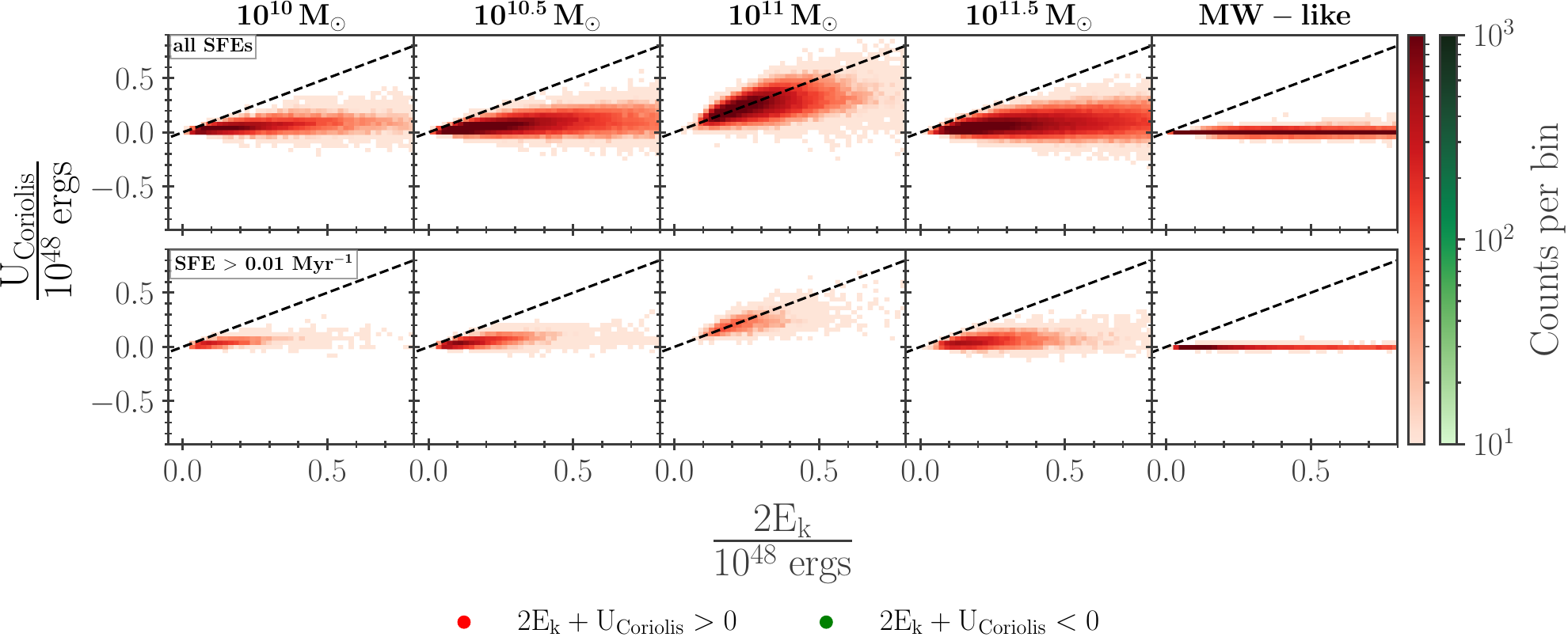}
    \caption{2-D histograms of two times the total kinetic energy versus the Coriolis energy term, in the style of Figure~\ref{fig:SG_vs_coriolis}. Pixels are colored based on the sum of the force terms, where reds represent a positive sum (expansion, or support against collapse) while greens are a negative sum (collapsing overdensity). The dashed black line represents a one-to-one relationship. As in Figure~\ref{fig:axi_anticorrelation}, the top row is all overdensities, while the bottom row is limited to overdensities with SFE $> 0.01 \; {\rm Myr^{-1}}$. All overdensities appear to be supported against collapse when only the contributions from the kinetic energy and Coriolis force are considered. However, the suppressed ETG ($\rm 10^{11}M_{\odot}$) does not maintain a distribution as flat as the other galaxies, where the one-to-one line shows a Coriolis energy term nearly equivalent in magnitude to $2E_{\rm k}$ .}
    \label{fig:KE_vs_coriolis}   
\end{centering}
\end{figure*}

Figure~\ref{fig:KE_vs_coriolis} is identical to Figure~\ref{fig:SG_vs_coriolis}, but now shows the kinetic energy term of the MVT, $2E_k$, instead of the self-gravity. As might be expected, all overdensities in this case appear to have a net positive term, indicating support against collapse. Distributions appear to be identical in shape independent of SFE for each galaxy, but we again note an important difference in the suppressed ETG: As the amount of kinetic energy increases, so does the Coriolis energy term, similar to what is seen with $\alpha_{\rm vir}$ in Figure~\ref{fig:alpha_vs_coriolis}. From the dashed one-to-one line in Figure~\ref{fig:KE_vs_coriolis}, the Coriolis energy contribution in the suppressed ETG is actually similar to that from $2E_{\rm k}$.

\begin{figure*}
\begin{centering}
    \includegraphics[width=\linewidth]{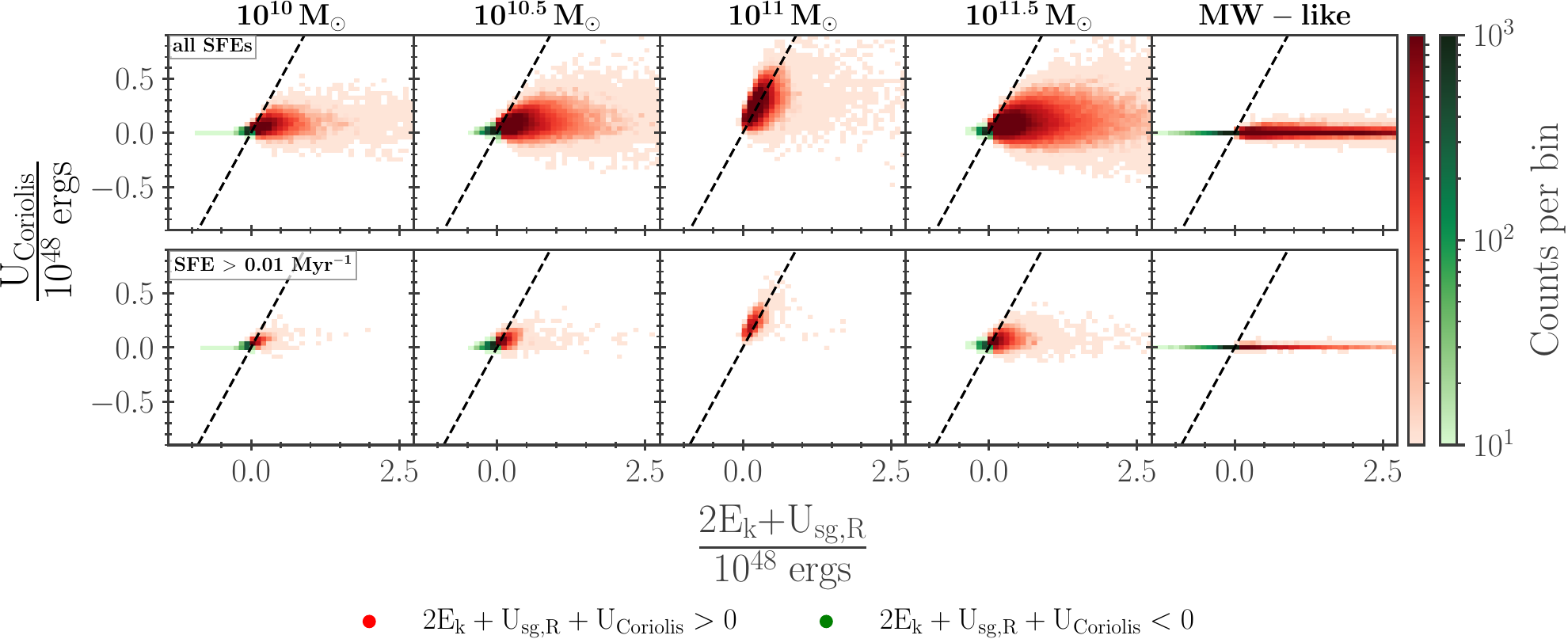}
     \caption{2-D histograms of two times the total kinetic energy plus the self-gravity in the midplane direction, versus the Coriolis energy term, in the style of Figure~\ref{fig:SG_vs_coriolis}. Pixels are colored based on the sum of the force terms, where reds represent a positive sum (expansion, or support against collapse) while greens are a negative sum (collapsing overdensity). The dashed black line represents a one-to-one relationship, where both components are equal. As in Figure~\ref{fig:axi_anticorrelation}, the top row is all overdensities, while the bottom row is limited to overdensities with SFE $> 0.01 \; {\rm Myr^{-1}}$. As in Figure~\ref{fig:KE_vs_coriolis}, all overdensities from the suppressed ETG ($\rm 10^{11}M_{\odot}$) appear to be supported against collapse, and a steep relationship between the sum of kinetic energy and self-gravity and the Coriolis energy term is visible, in contrast with the other ETGs.}
    \label{fig:KEplusSG_vs_coriolis}   
\end{centering}
\end{figure*}

Finally, Figure~\ref{fig:KEplusSG_vs_coriolis} combines Figures~\ref{fig:SG_vs_coriolis} and \ref{fig:KE_vs_coriolis} and now shows all relevant terms for the modified virial theorem. In the top panel, galaxies show only a few percent of identified overdensities as being supported against collapse, while this increases to 20-30\% in the bottom panels, where the SFE is higher (with the exception of the suppressed ETG). This might be expected due to the reasoning presented earlier; many of the star-forming gas cells have reached temperature and density conditions that allow for star formation to occur, but do not represent true GMC conditions, particularly in the ETGs.

However, while most galaxies show some amount of net compressive overdensities (green pixels), little-to-none are visible in the suppressed ETG (middle column). Again, this galaxy experiences a much steeper relationship between the Coriolis energy and the energy sum, $2E_{\rm k}+U_{\rm sg,R}$, which from Figure~\ref{fig:KE_vs_coriolis} can be interpreted to be due to the dependence on the kinetic energy term. This relationship is not visible in the self-gravity (Figure~\ref{fig:SG_vs_coriolis}).

\subsection{What Drives the Coriolis Force?}

It has become clear that in the reference frame of the cloud, the size of the Coriolis force term is the key differentiator between early-type galaxies with different levels of star formation. For a complete physical understanding of this scenario, we now explore what drives the Coriolis force itself. 

The calculation for the energy from the Coriolis term, given in Equation~\ref{eqn::MVT9_Rextnax}, can be simplified by defining the internal angular momentum of the cloud (in the $z-$direction; also known as `spin') such that $-L_{z,i}=m_i(\dot{d}_{x,i}d_{y,i}-\dot{d}_{y,i}d_{x,i})$. With this, Equation~\ref{eqn::MVT9_Rextnax} then simply becomes

\begin{equation}
    U_{\rm Coriolis} = \sum_i 2 \,\Omega_0\,w_iL_{z,i}.
\end{equation}

Neglecting the weighted kernel $w_i$, the Coriolis energy term is then only dependent on $\Omega_0$, the angular velocity of the overdensity's COM around the galactic center, and $L_{z,i}$, the overdensity's spin angular momentum. We can now investigate these two terms to determine if one appears to be the driving factor in star formation suppression. 

\begin{figure*}
\begin{centering}
    \includegraphics[width=\linewidth]{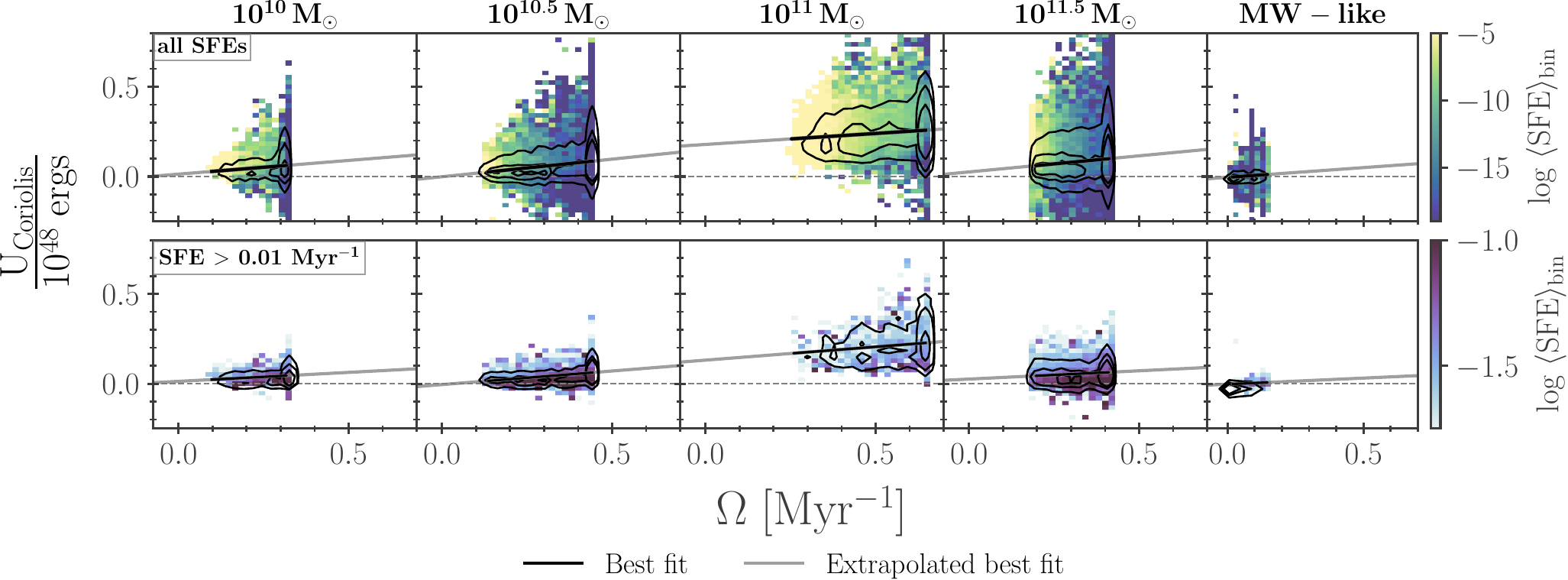}
    \caption{2-D histograms of the overdensity angular velocity (relative to the galactic center), $\Omega$, versus the energy term from the Coriolis force, colored by the mean SFE in each bin. Unfilled contours show the 50\%, 75\%, and 95\% data inclusion regions, while the solid line is the least squares regression. Dashed grey line is where the Coriolis energy term is 0, while the black solid line is the linear least squares regression. Grey solid line is an extrapolation of this linear best fit. As in Figure~\ref{fig:axi_anticorrelation}, the top row is all overdensities, while the bottom row is limited to overdensities with SFE $> 0.01 \; {\rm Myr^{-1}}$. All ETGs display a similar relationship between $\Omega$ and $U_{\rm Coriolis}$ with an exception of the suppressed ETG.}
    \label{fig:Omega_vs_coriolis}   
\end{centering}
\end{figure*}

We show 2-D distributions of $\Omega$ versus $U_{\rm Coriolis}$ in Figure~\ref{fig:Omega_vs_coriolis}, which are identical to the style of Figure~\ref{fig:alpha_vs_coriolis}. However, instead of being colored by mean bin SFR, pixels are now colored by the mean bin SFE (we note that the distribution is identical whether colored by SFR or SFE). Values of $\Omega$ are calculated by interpolating the distribution of $\Omega(R)$ in Figure 3 of \citetalias{Jeffreson2024b}, where $R$ is the overdensity's distance from the galactic center. This explains the sharp minima and maxima seen for $\Omega$, as we limit the analysis range to $R_{\rm min}=0.3$ kpc and an $R_{\rm max}=1.5$ kpc for ETGs, and $R_{\rm max}=13$ kpc for the MW-like galaxy.

In the ETGs, we generally see a similar positive slope, but with both an increase in the range and in the values of $\Omega$. The suppressed ETG, however, has both the largest range of $\Omega$ and the largest values, indicating more rotational support.

In all galaxies, the bottom panels (SFE$>0.01 \rm \; Myr^{-1}$) show that higher SFEs are found, regardless of $\Omega$, at lower $U_{\rm Coriolis}$.

\begin{figure*}
\begin{centering}
    \includegraphics[width=\linewidth]{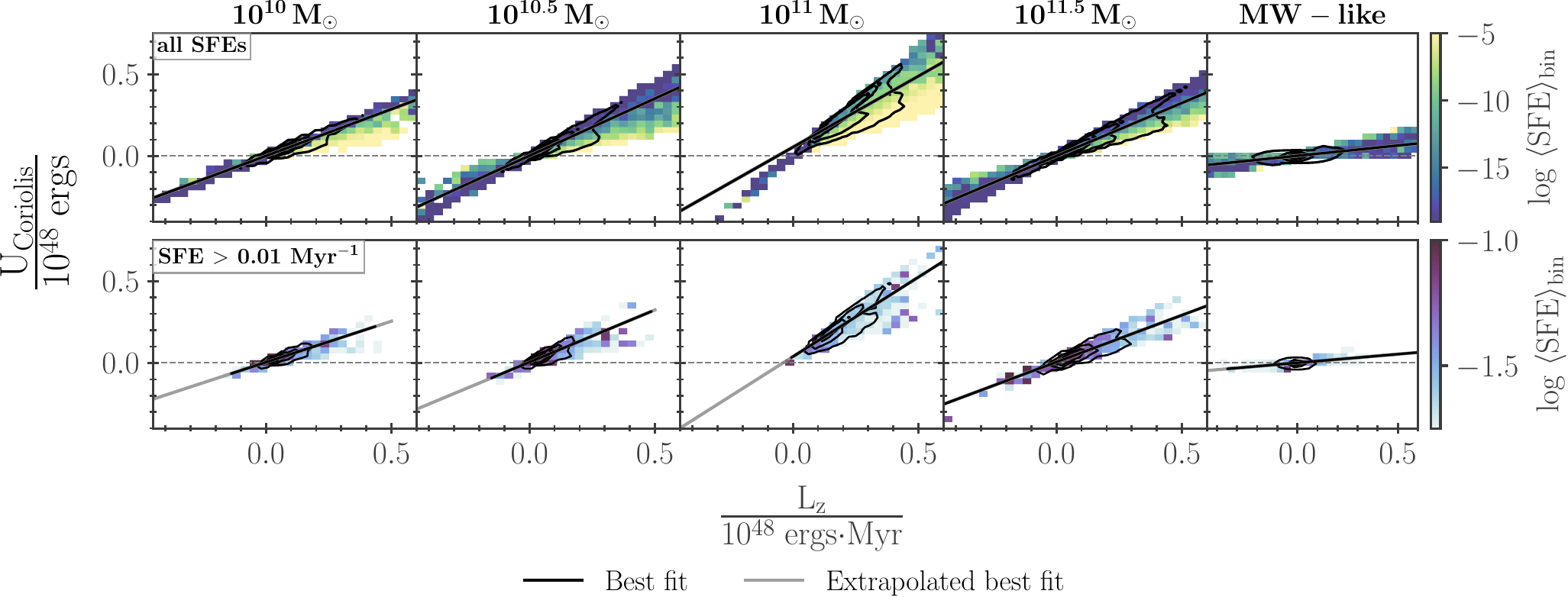}
    \caption{2-D histograms of the overdensity's spin angular momentum, $\rm L_z$, versus the energy term from the Coriolis force, colored by the mean SFE in each bin. Unfilled contours show the 50\%, 75\%, and 95\% data inclusion regions, while the solid line is the least squares regression. Dashed grey line is where the Coriolis energy term is 0, while the black solid line is the linear least squares regression. Grey solid line is an extrapolation of this linear best fit. As in Figure~\ref{fig:axi_anticorrelation}, the top row is all overdensities, while the bottom row is limited to overdensities with SFE $> 0.01 \; {\rm Myr^{-1}}$. The MW-like galaxy shows largely symmetrical $L_z$ concentrated around $L_z=0$, while the ETGs have a larger distribution of $L_z>0$. This effect is more pronounced in the suppressed ETG.}
    \label{fig:Lz_vs_coriolis}  
\end{centering}
\end{figure*}

It is also instructive to examine $L_z$, another component which may illustrate the strength of the Coriolis force in the overdensity reference frame. The relationships between $L_z$ and $U_{\rm Coriolis}$ are visible in Figure~\ref{fig:Lz_vs_coriolis}. Here, the MW-like galaxy appears to have a largely symmetrical distribution that is relatively flat, with most of the data concentrated around $U_{\rm Coriolis}$ and $L_z$ of 0. The ETGs show stronger positive slopes of $L_z$ as opposed to largely symmetrical distributions, and again the suppressed ETG shows a stronger slope and higher values of $L_z$. 

In the regime of high SFE (bottom rows of Figure~\ref{fig:Lz_vs_coriolis}), the largest mean SFEs are concentrated around $L_z=0$. In fact, most of the ETGs show distributions centered around around $L_z=0$ albeit with some positive scatter. The suppressed ETG is the only galaxy to not follow this behavior, instead having an almost entirely positive internal angular momentum, or spin. 

\begin{figure*}
\begin{centering}
    \includegraphics[width=\linewidth]{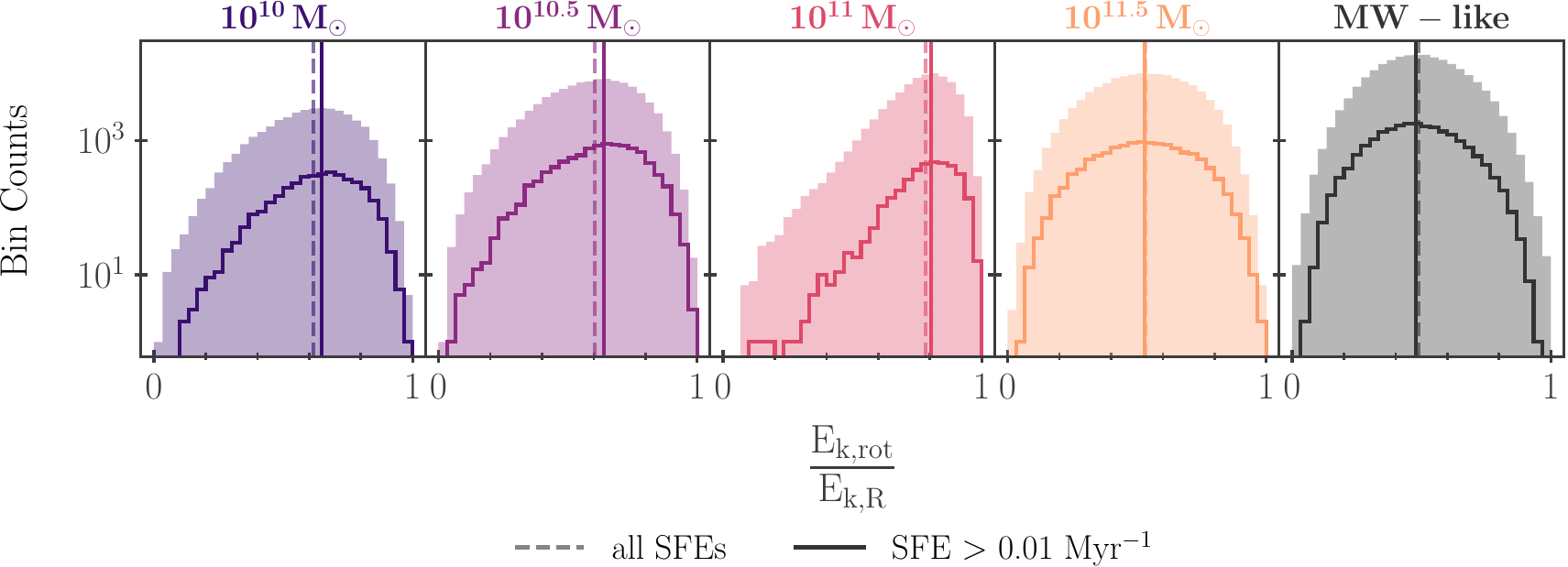}
    \caption{Distributions of rotational kinetic energy over total kinetic energy (in the midplane direction) across overdensities for each galaxy (columns) from Equation~\ref{eqn::MVT5_rotational_kineticenergy}. Histogram styles and vertical lines (medians) are as Figure~\ref{fig:all_force_hist}, with lighter filled histograms and dashed line being for all overdensities, and unfilled histograms and solid line for overdensities with SFE $\rm > 0.01 \; Myr^{-1}$. All galaxies, including the MW-like, show symmetrical distributions with medians around 0.5-0.6. The suppressed ETG, however, shows a much larger value for the rotational kinetic energy, with $\rm E_{k,rot}/E_{k,R}\approx0.8$}
    \label{fig:fraction_KErot_KEtotal}   
\end{centering}
\end{figure*}

Both the angular velocity and spin of the overdensities in the suppressed ETG are elevated in line with the elevated Coriolis energy term. Physically, we suggest that the higher degree of galactic rotation in this galaxy creates a torque that supports overdensities against collapse. Parcels of gas moving towards the Galactic center, for example, are deflected in the direction of the galaxy's rotation. This increases the spin ($L_z$) of the cloud and can further promote the effects of shear, dispersing the cloud or providing rotational support against collapse. 

In turn, the rotational kinetic energy of the cloud is elevated. While we acknowledge that little differences are found in the magnitude of the kinetic energies from Figure~\ref{fig:all_force_hist}, this is because these terms account for energies in all directions ($R$ and $z$). Instead, if we investigate the fraction of rotational kinetic energy compared to the midplane kinetic energy ($E_{\rm k,rot}/E_{\rm k,R}$), as shown in Figure~\ref{fig:fraction_KErot_KEtotal}, we indeed see an elevated fraction of rotational kinetic energy in the suppressed ETG, consistent with the physical scenario outlined above. While most galaxies see median values of $E_{\rm k,rot}/E_{\rm k,R}\approx0.5-0.6$, the $10^{11} \, \rm M_\odot$ ETG shows a slightly higher median. 

Therefore, even though the GalactISM simulation suite does not directly account for the additional term of the Coriolis force in the virial parameter and subsequent star formation prescription, the impact is seen through the contribution of rotation to the kinetic energy and resulting suppressed star formation.

\section{Discussion}\label{Sec::Discussion}

Having presented the overdensity properties, energy terms, and the related driving mechanisms, we now contextualize our results and compare to previous literature. We first review how star formation suppression has been previously discussed and compare with this paper (\ref{Sec::Discussion_PriorWork}), then connect to an observational point of view (\ref{Sec::Discussion_Observations}), and review the caveats of our analysis (\ref{Sec::Discussion_Caveats}).

\subsection{Comparison to Prior Simulation Works}\label{Sec::Discussion_PriorWork}

Using a modified virial theorem to study cloud dynamics is not necessarily a new idea \citep{Meidt2018, Meidt2020, Liu2021}. In addition to \citet{Liu2021}, who use this in an observational study of GMCs, \citet{Meidt2018} and \citet{Meidt2020} apply this analytically and develop predictions for observational signatures. Our results agree well with the conclusion by \citet{Meidt2020} that the Coriolis force should not be negligible on cloud scales, and can have the same magnitude as that of self-gravity. In fact, from Figure~\ref{fig:SG_vs_coriolis}, we suggest that the Coriolis force may be even larger than self-gravity for compact ETGs with strong rotation. More broadly, we note the importance of studying star formation suppression using these new methods.

Since the introduction of `morphological' quenching by \citet{Martig2009}, suppressed star formation in ETGs is often associated with increased classical Toomre-$Q$ parameters, velocity dispersion, galactic rotation/shear, and S\'ersic indices, where an increase in the latter parameter usually corresponds to a more spherical stellar luminosity profile \citep{Martig2009}. In particular, we note that previous arguments (particularly those made with Toomre-$Q$) make several limiting assumptions about the ISM, including being single-phase, infinitesimally thin gas disks, and constant star formation efficiencies \citep{Murray2011a, Martig2013, Gensior2020, Kretschmer2020}. 

These simplistic assumptions are often justified because previous simulation works lack the resolution to accurately capture a multiphase ISM \citep{Martig2009, Martig2013, Gensior2020}. When a multiphase ISM is resolved and varying SFEs are considered, the classical Toomre-$Q$ parameter is no longer a clear criterion for star formation suppression. Instead, on cloud scales, the modified virial theorem allows for the impact of the external gravitational potential and rotation to be considered while also allowing for realistic assumptions about the ISM. 

For example, a direct predecessor to this paper, \citet{Gensior2020} combines the powerful approach of testing both the effect of subgrid star formation models and the direct impact of morphology. They perform hydrodynamical simulations of isolated galaxies, varying the stellar bulge mass and scale radius and keeping initial stellar, halo, and gas masses constant ($f_{\rm gas}\approx5\%$). Each model uses a constant star formation efficiency, as also tested by \citet{Kretschmer2020}, as well as the dynamics-dependent model with parameterizations by \citet{Padoan2017} used here. While they find that models with strong central stellar components always increase the velocity dispersion and stability of the gas disk near the galactic center, this is more amplified in the dynamical SF simulations than those with constant SFE. 

In this paper, all ETGs have been initialized such that the disk-to-bulge mass ratio is a constant 0.2, but our suppressed ETG at $10^{11}\,{\rm M_{\odot}}$ shows the highest bulge concentration with a central stellar surface density of $\mu_\star=1700\,M_{\odot}\,{\rm pc^{-2}}$ (measured within the stellar half-light radius). As \citetalias{Jeffreson2024b} notes, this central concentration is 70\% larger than the ETG with the second most compact bulge, and a correlation is found between $\mu_\star$ and galactic rotation across the entire simulation suite.

Furthermore, from Figure~\ref{fig:overdensity_props} (see also Figure 7 of \citetalias{Jeffreson2024b}) and in contrast with the mentioned literature on star formation suppression, the suppressed ETG does not show significantly elevated velocity dispersions or Toomre-$Q$. While \citetalias{Jeffreson2024b} presents this galaxy as having the smoothest gas disk with fewer voids, the Toomre-$Q$ parameter is similar (and high) in all of the ETGs, and this galaxy shows the lowest gas radial velocity dispersion. Because of the above mentioned assumptions regarding high $Q$ being linked to dynamical suppression, we reiterate that this may only be true in an unresolved ISM. Instead, the contribution of the external gravitational potential and internal overdensity rotation may be more accurate tracers of star formation suppression when multiple gas phases are resolved.

\subsection{Observational Connection}\label{Sec::Discussion_Observations}

While many of the parameters here are difficult to derive observationally, especially with the primary observable properties of GMCs being their size, velocity dispersion, and luminosity \citep{Larson1981, Solomon1987}, recent surveys have shown the ability to resolve individual GMCs within early-type galaxies in the local Universe \citep{Utomo2015, Liu2021, Williams2023, Lu2024, Lu2025}. Furthermore, many of these observed GMCs are thought to be directly impacted by both galactic shear and internal rotation. 

Through $^{12}\rm{CO}(2-1)$ observations of NGC 4526, \citet{Utomo2015} present the first observational GMC catalog of an ETG, consisting of nearly a hundred clouds and their properties. With such high resolution observations ($\sim$20 pc), they obtain cloud kinematics by comparing the alignment of smaller-scale velocity gradients to the galaxy velocity field, showing that GMC rotation is driven by the galaxy's shear (and corresponding potential). Furthermore, they find that clouds near the galactic center are unlikely to overcome these rotational forces.

\citet{Liu2021} use even higher-resolution ($\sim$12 pc) observations of NGC 4429 to identify about twice as many GMCs as presented in the first catalog by \citet{Utomo2015}. These GMCs also shown to possess a large amount of internal rotation that is driven by the galaxy's potential, and after removing contributions from this rotation, the internal virial parameter decreases and virial equilibrium can be established. In addition, \citet{Liu2021} consider the importance of the external gravitational contribution (including shear), using a modified virial theorem, demonstrating the observational relevance of such a method. As a result, many of the clouds in NGC 4429 appear to experience the effects of tidal forces and shear by being radially elongated.

\subsection{Caveats}\label{Sec::Discussion_Caveats}

We find it important to elaborate on a few specific caveats of this paper. 

First, because many of the properties analyzed here may be dependent on the size of the cloud (overdensity) analyzed, we find it important to rationalize our choice of scale and show how results are robust even when this scale is chosen to be uniform. 

As described in Section~\ref{Sec::Physics_SF}, the scale of each simulation overdensity, over which the virial parameter and subsequent SFE are calculated, is determined by the Sobolev length $\ell = \rho/|\nabla \rho|$. Because this is dependent on the local ISM conditions, each overdensity of every galaxy therefore has a unique $\ell$, visible in the top row of Figure~\ref{fig:overdensity_props}.  While \citet{Gensior2020} shows that this calculation is only weakly dependent on resolution and allows for a more dynamics-dependent star formation model, we acknowledge that terms in this paper may be affected by such a scale definition. 

\begin{figure*}
\begin{centering}
    \includegraphics[width=\linewidth]{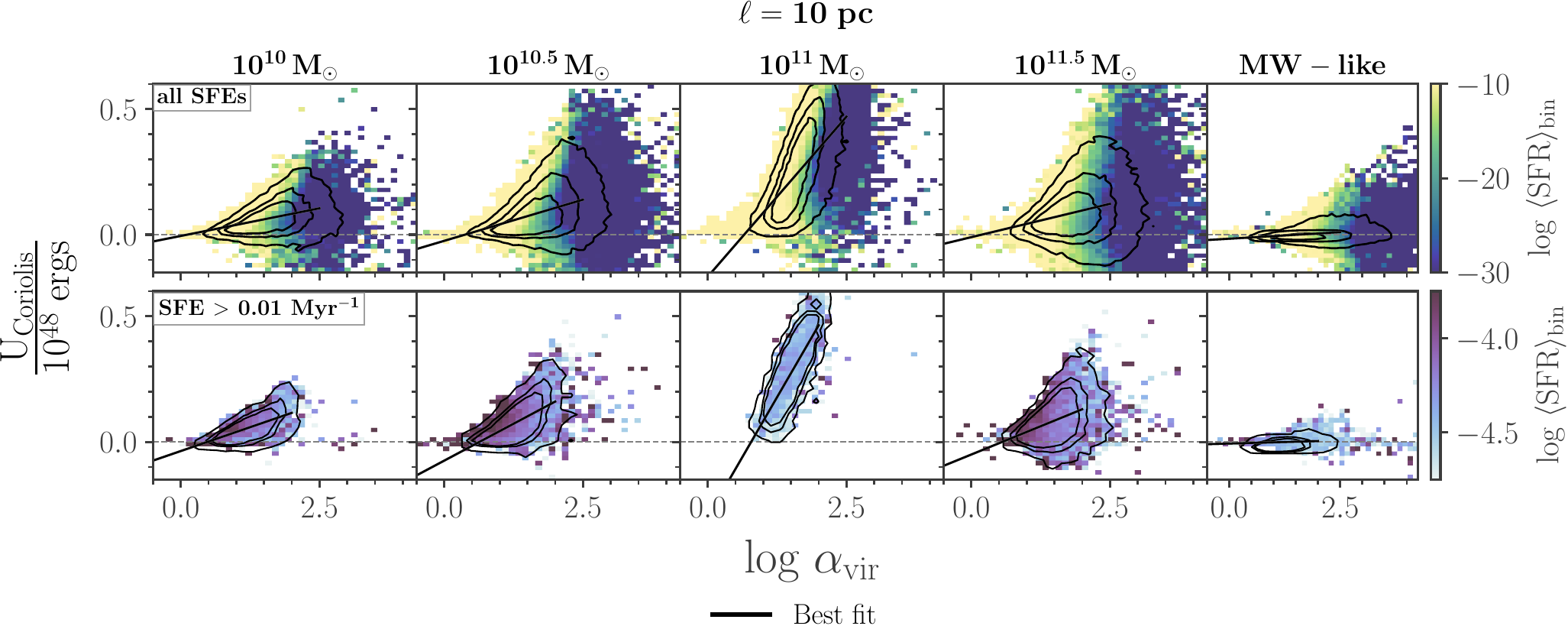}
    \caption{2-D histograms of the classical virial parameter $\alpha_{\rm vir}$ versus the energy term from the Coriolis force, colored by the mean SFE in each bin, for an overdensity radius of 10 pc instead of the Sobolev length $\ell=\rho/|\nabla\rho|$.  Unfilled contours show the 50\%, 75\%, and 95\% data inclusion regions, while the solid line is the least squares regression. Dashed grey line is where the Coriolis energy term is 0, while the black solid line is the linear least squares regression. As in Figure~\ref{fig:axi_anticorrelation}, the top row is all overdensities, while the bottom row is limited to overdensities with SFE $> 0.01 \; {\rm Myr^{-1}}$. Despite the fixed scale, results are identical to Figure~\ref{fig:alpha_vs_coriolis}. }
    \label{fig:10pc_alpha_vs_coriolis}   
\end{centering}
\end{figure*}

To show that our primary findings are robust when a uniform scale choice is applied to all overdensities of all galaxies, we repeat all previously-shown calculations for $\ell=10$ pc. This fixed distance is approximately the median radius of all galaxy overdensities, and similar to the scale of the Toomre wavelength from \citetalias{Jeffreson2024b}. We essentially retain the same number of `clouds', which are defined from star-forming gas cells, but they now include all gas cells (from which energy terms are calculated) within a distance of 10 pc. For comparison with the Sobolev length, we do not change the smoothing length $h$ used in the calculation, which is defined by \citet{Gensior2020} to be twice the distance of the 32nd nearest weighted gas cell. 

Figure~\ref{fig:10pc_alpha_vs_coriolis} shows distributions from Figure~\ref{fig:alpha_vs_coriolis} in this $\ell=10$ pc calculation, consisting of the virial parameter, which captures the self-gravity and kinetic energy terms, the Coriolis energy term, and the mean SFR per pixel, effectively capturing all terms of the MVT that we do not rationalize discarding. Some minor differences can be seen at the uniform scale, such as lower $\alpha_{\rm vir}$ in some galaxies and $U_{\rm Coriolis}$ decreasing for the suppressed ETG and increasing for the MW (following the scale changes), but the overall relationships do not change. Distributions in the unsuppressed ETGs are similar, the MW-like galaxy keeps a flat profile, and the suppressed ETG shows higher $U_{\rm Coriolis}$ with a steeper dependence on $\alpha_{\rm vir}$ and lower SFRs (bottom row).

Second, during the calculation of the Sobolev length $\ell$, \citet{Gensior2020} notes that this occurs over an iterative tree-walk until a defined convergence criteria is met. While we do not recount the details of this calculation here, and instead point the reader to their Section 2.2, we do note that the convergence criteria results in a slight numerical instability between the tree-walk length scale, $\ell_{\rm tw}$, and the Sobolev length, $\ell=\rho/|\nabla\rho|$. All parameters computed here are done so using the Sobolev length, while the exact radius of clouds is $\ell_{\rm tw}$. However, this offset between $\ell$ and $\ell_{\rm tw}$ is always within the allowed convergence criteria set by the algorithm, which we find here to be no more than about 1-2 pc.

In addition, many of the overdensity properties here, in contrast with \citetalias{Jeffreson2024b}, are computed after the simulation runtime and not over hierarchical timesteps; these are instantaneous values. In addition, this snapshot is the final result in the simulation suite, stated in Section~\ref{Sec::Sims_InitialConditions} to be $t_{\rm end}= 400$ Myr for the ETGs and $t_{\rm end}=600$ Myr for the MW-like galaxy.

Finally, when calculating the gravitational potential $\Phi$, we employ an interpolation-based method of varying resolution to calculate the effect from gas particles $\Phi_{\rm gas}$, dark matter particles $\Phi_{\rm DM}$, the stellar disk $\Phi_{\star,\rm disk}$, and the stellar bulge $\Phi_{\rm \star, bulge}$. Described in Appendix B of \citetalias{Jeffreson2024b}, the potential is calculated in bins of $R, \theta,z$ by interpolating values across the nearest 150 particle centroids (of any type: gas, stellar, or dark matter) to the bin center from radial basis function interpolation \citep{Hines2023}. Because this method is not exact, the interpolation introduces possible noise and sources of error. However, we note that computing the potential based on all corresponding particles is computationally expensive, and little difference in seen when calculating the potential over different bin sizes.

\section{Conclusions and Future Work} \label{Sec::conclusions}

In this work, we use a modified virial theorem (MVT) to investigate the driving force behind the suppression of star formation in an early-type galaxy from the suite of Milky Way-like and early-type galaxies presented by \citetalias{Jeffreson2024b}. We investigate the star-forming overdensities and their properties across the five simulations, and calculate the energy contributions from the kinetic energy, self-gravity, external gravitational forces, and the Coriolis force. To link distinct energy terms to the star formation prescription, we further investigate relationships with the virial parameter $\alpha_{\rm vir}$, galactic angular velocity $\Omega$ at the position of each overdensity, and internal angular momentum (`spin') $L_z$ relative to the cloud center of mass. 

Our primary findings are as follows:
\begin{enumerate}
    \item In comparison with the other galaxies in the suite, the suppressed ETG has lower levels of self-gravity and elevated Coriolis forces acting on its overdensities, due to the steeper external gravitational potential gradient $\partial \Phi_{\rm ext}/\partial R$ of the host galaxy. These increased Coriolis forces are associated with increased rotational kinetic energy that provides support against gravitational collapse.
    \item Although this gravitational support is associated with a higher degree of galactic rotation $\Omega$, in contrast with previous works on star formation suppression, it is not strongly reflected in the Toomre-$Q$ parameter. The modified virial theorem may provide a better indication of the onset of star formation suppression on cloud scales than the classic Toomre-$Q$ parameter.
    \item Our results are in agreement with previous studies demonstrating an association between star formation suppression and the central concentration of galactic stellar mass, which produces a higher rate of galactic rotation in the gas disk, and thus enhances the Coriolis term in the modified virial theorem.
    \item All other galaxies, whether early-type or MW-like, have overdensities that are governed predominantly by kinetic energy and self-gravity alone, with negligible contributions from the external gravitational potential. These overdensities are therefore well described by the traditional form of the virial theorem.
\end{enumerate}

By framing star formation through the lens of a modified virial theorem, these results highlight the importance of the interaction between galactic shear and cloud-scale gas dynamics in regulating star formation. In particular, the elevated Coriolis forces and rotational support observed in the suppressed ETG suggest that internal dynamical processes, rather than solely environmental effects (such as mergers or black holes) can play a decisive role in lowering star formation efficiencies. 

Future investigations using more centrally-compact simulated galaxies at cloud-scale resolutions will serve to further constrain the relationship between rotational support, galactic morphology, and star formation rate. On the observational side, works such as \citet{Liu2021, Lu2024, Lu2025} are now beginning to resolve significant populations of star-forming overdensities in rotationally-extreme environments, including the internal gas velocities of these overdensities (see Section~\ref{Sec::Discussion_Observations} and references therein). In the future, surveys such as WISDOM may provide a means to more closely observationally corroborate the conclusions of this work across galaxy populations.

\begin{acknowledgments}
L.E.P. acknowledges support from the NSF GRFP. S.M.R.J. is supported by the Simons Foundation through the Learning the Universe collaboration. GLB acknowledges support from the NSF (AST-2108470, AST-2307419), NASA TCAN award 80NSSC21K1053, and the Simons Foundation through the Learning the Universe Collaboration.
We thank Volker Springel for providing us access to {\sc AREPO}.
\end{acknowledgments}

\appendix

\section{Axisymmetric Midplane Components}\label{appendix::Rextaxi_components}

\begin{figure*}
\begin{centering}
    \includegraphics[width=\linewidth]{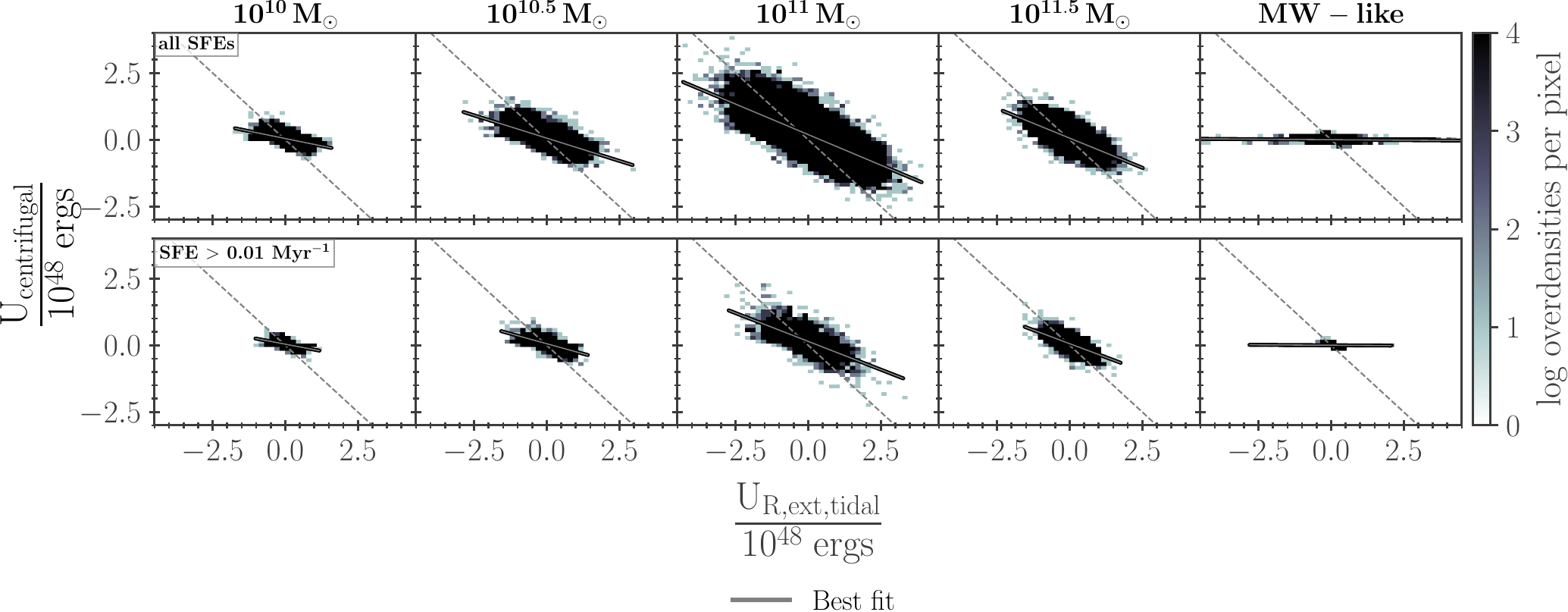}
    \caption{2-D histograms of energy from both terms in the axisymmetric midplane component from Equation~\ref{eqn::MVT8_Rextax}: the tidal force (x-axis) and centrifugal force (y-axis). The top row shows the distribution across all overdensities, while the bottom row is limited to overdensities with SFE $> 0.01 \; {\rm Myr^{-1}}$. Light grey dashed line shows where both terms are equal, and the darker solid line is the linear least squares regression.}
    \label{fig:axi_anticorrelation}   
\end{centering}
\end{figure*}

While $U_{\rm R,ext,ax}$ is actually comprised of two terms in Equation~\ref{eqn::MVT8_Rextax}, the `tidal' force $U_{\rm R,ext,tidal}$ and the centrifugal force $U_{\rm centrifugal}$, we group these energy contributions into one because they are not mutually exclusive---both are dependent on the gravitational potential. However, it is also logical to expect that both terms should cancel in an axisymmetric steady-state gas disk. 

If we consider an axisymmetric external potential $\Phi_{\rm ext}(R)$, the partial derivatives of Equation~\ref{eqn::MVT8_Rextax} become $\frac{x}{R}\frac{\partial \Phi_{\rm ext}}{\partial R}$ and $\frac{y}{R}\frac{\partial \Phi_{\rm ext}}{\partial R}$, respectively. Then, for circular rotation at some radius $R_0$, the partial derivative with respect to $R$ is 
\begin{equation}
    \frac{\partial \Phi_{\rm ext}}{\partial R}|_{R_0}=R_0\Omega_0^2.
\end{equation}
If we substitute this into the tidal term of Equation~\ref{eqn::MVT8_Rextax}, it results in 
\begin{equation}
    -\frac{\partial \Phi_{\rm ext}}{\partial R}\frac{xd_x + yd_y}{R_0}=-\Omega_0^2(xd_x+yd_y),
\end{equation}
which cancels exactly with the centrifugal contribution. Where the tidal field would compress radially, the differential centrifugal force would expand, and vice versa.

For an axisymmetric potential, a cloud's COM is located at some position $R_0$ with a circular orbit in the midplane. By definition of the circular orbit and calculation for $\Omega$, the radial gravitational force from the potential on the cloud's COM is exactly balanced by the centrifugal force such that $\frac{\partial \Phi}{\partial R}|_{R_0}=R_0\Omega^2$. Therefore, at the COM, $U_{\rm R,ext,ax}$ exactly cancels.

However, when we consider the displacement ($d_{x,i}, d_{y,i}$) of other gas cells and expand the potential about the COM in the noninertial reference frame, the calculation takes on additional higher-order terms (see Equation 6 in \citealt{Meidt2018}). Furthermore, in a realistic gas disk, this term may not show a distribution centered exactly around $U_{\rm R,ext,ax}=0$ due to imperfect axisymmetry, which may be caused by the presence of spiral arms, bars, or other features. 

Figure~\ref{fig:axi_anticorrelation} shows the relationship between our calculated $U_{\rm centrifugal}$ and $U_{\rm R,ext,tidal}$, where the dark solid line is the linear least squares regression (line of best fit), and the lighter dashed line is an exact anti-correlation. While distributions are identical across all overdensities (top row) and those with higher SFEs (bottom row), we note a clear difference between the ETGs and MW-like galaxy. While the ETGs all maintain a constant slope of approximately -0.5, the MW-like galaxy's energy from the centrifugal force is almost constantly flat with increasing tidal forces. The distribution is almost exactly symmetric about $U_{\rm R,ext,tidal}=0$. 

In all galaxies, the sum of the two terms (as represented in Figure~\ref{fig:all_force_hist}) is nearly zero with the exception of the suppressed ETG, though this distribution is still symmetric.

We also note that some noise in the calculations are to be expected due to the calculation of the gravitational potential, which we discuss further in Section~\ref{Sec::Discussion_Caveats}. For these reasons, similar to previous literature implementing a modified virial theorem \citep{Meidt2018}, we choose to neglect the inclusion of $U_{\rm R,ext,ax}$ for most of this paper. We note that this does not affect any presented conclusions.

\section{Vertical Components}\label{appendix::vertical_components}

\begin{figure*}
\begin{centering}
    \includegraphics[width=\linewidth]{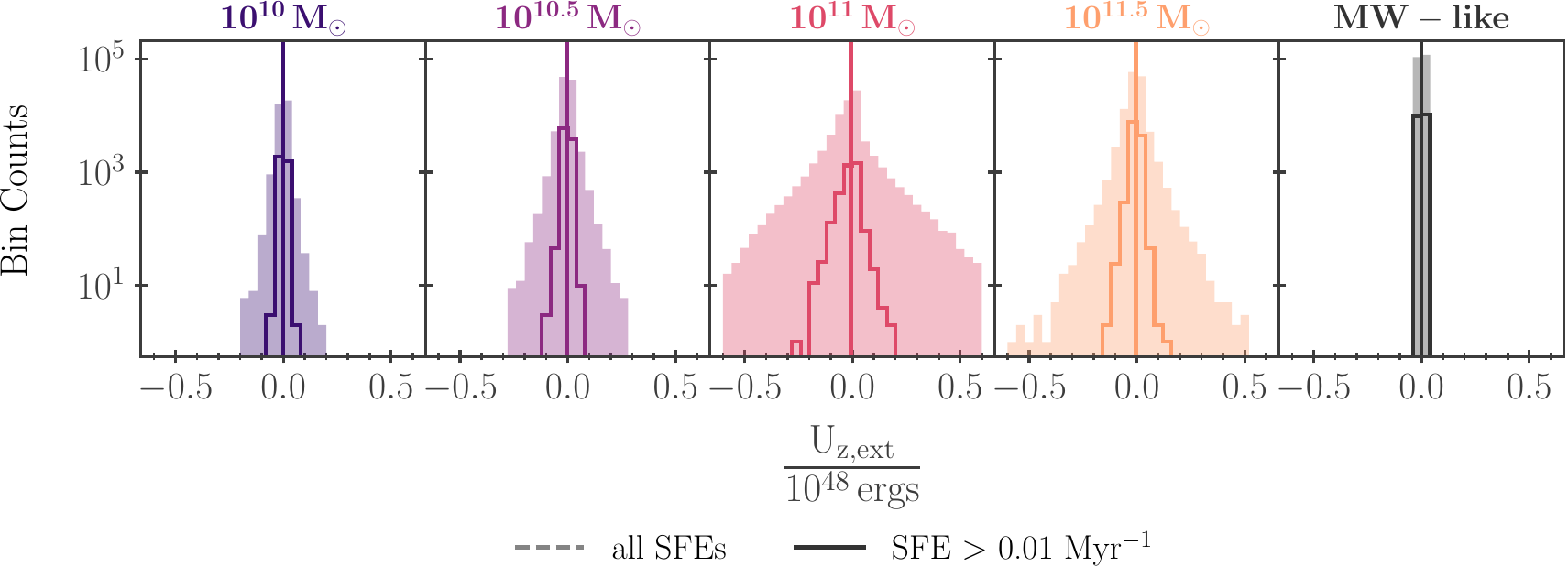}
    \caption{Distributions of energy from the external gravitational potential, perpendicular to the midplane, across overdensities for each galaxy (columns) from Equation~\ref{eqn::MVT7_zext}. Histogram styles and vertical lines (medians) are as Figure~\ref{fig:all_force_hist}, with lighter filled histograms and dashed line being for all overdensities, and unfilled histograms and solid line for overdensities with SFE $\rm > 0.01 \; Myr^{-1}$. All galaxies have largely symmetric distributions with medians of approximately $U_{\rm z,ext}=0$. The MW-like galaxy experiences the least contribution from the external gravitational potential, likely because it exhibits the thickest gas disk in Figure~\ref{fig:gal_imshows}.}
    \label{fig:zext_dist}   
\end{centering}
\end{figure*}

 $U_{\rm z,ext}$, the energy contribution from the external gravitational potential perpendicular to the midplane, is defined by Equation~\ref{eqn::MVT7_zext} and has distributions shown in Figure~\ref{fig:zext_dist}. 

As mentioned in Section~\ref{Sec::vertical_forces}, every distribution is nearly symmetrical with a median around 0, though the ETGs experience a slight vertical skew, especially in the $10^{11}\,\rm M_{\odot}$ galaxy. 

This force is likely so low and negligible to our analysis due to the thin nature of the gas disks from Figure~\ref{fig:gal_imshows}. In particular, we point out that the gas disk with the lowest scale height, the suppressed ETG, has the widest range in values of $U_{\rm z,ext}$, extending beyond $\pm\,5\times10^{47}$ ergs in all overdensities (filled histogram). In contrast, the MW-like galaxy with the thickest gas disk experiences a much narrower range with values extending to $\pm\,3\times10^{46}$.


\bibliography{ref}{}
\bibliographystyle{aasjournal}



\end{document}